\documentclass[reqno,a4paper,11pt]{article}
\pdfoutput=1
\usepackage{xcolor}

\usepackage{graphicx}
\usepackage[textwidth = 430 pt, textheight = 630 pt]{geometry}

\definecolor{MyDarkBlue}{rgb}{0.15,0.25,0.45}
\usepackage{epsfig,rotating}
\usepackage{amsmath,amssymb}
\usepackage{amsfonts}
\usepackage{mathrsfs}
\usepackage{bbm}
\usepackage[normalem]{ulem}

\usepackage{latexsym}
\usepackage{amsthm}
\usepackage[all,knot]{xy}
\xyoption{arc}

\usepackage[utf8x]{inputenc}

\usepackage{hyperref}
\hypersetup{
hypertexnames=false,
colorlinks=true,
citecolor=MyDarkBlue,
linkcolor=MyDarkBlue,
urlcolor=MyDarkBlue,
pdfauthor={Patricia Ritter and Christian S\"amann},
pdftitle={L-infinity-Algebra Models and Higher Chern-Simons Theories},
pdfsubject={hep-th math-ph}
breaklinks=true
}

\usepackage{tikz}
\usepackage{mathtools}

\let\fn\footnote
\renewcommand{\footnote}[1]{\linespread{1.1}\fn{#1}\linespread{1.29}}

\makeatletter\renewcommand{\section}{\@startsection
{section}{1}{\z@}{-3.5ex plus -1ex minus
    -.2ex}{2.3ex plus .2ex}{\bf }}
\makeatletter\renewcommand{\subsection}{\@startsection{subsection}{2}{\z@}{-3.25ex
plus -1ex minus
   -.2ex}{1.5ex plus .2ex}{\bf }}
\makeatletter\renewcommand{\subsubsection}{\@startsection{subsubsection}{3}{-2.45ex}{-3.25ex
plus -1ex minus -.2ex}{1.5ex plus .2ex}{\it }}
\renewcommand{\thesection}{\arabic{section}}
\renewcommand{\thesubsection}{\arabic{section}.\arabic{subsection}}
\renewcommand{\@seccntformat}[1]{\@nameuse{the#1}.~~}

\renewcommand{\theequation}{\thesection.\arabic{equation}}
\makeatletter \@addtoreset{equation}{section}

\usepackage[toc,page]{appendix}

\newtheorem{thm}{Theorem}[section]
\renewcommand{\thethm}{\thesection.\arabic{thm}}

\newtheorem{remark}[thm]{Remark}

\newcommand{\myxymatrix}[1]{\vcenter{\vbox{\xymatrix{#1}}}}

\renewcommand{\appendices}{
\section*{Appendix}\label{appendices}\setcounter{subsection}{0}
\addcontentsline{toc}{section}{Appendix}
\setcounter{equation}{0}
\makeatletter
\renewcommand{\theequation}{\Alph{subsection}.\arabic{equation}}
\renewcommand{\thesubsection}{\Alph{subsection}}
\renewcommand{\thethm}{\Alph{subsection}.\arabic{thm}}
\@addtoreset{equation}{subsection}
\@addtoreset{thm}{subsection}
\makeatother
}

\def\slasha#1{\setbox0=\hbox{$#1$}#1\hskip-\wd0\hbox to\wd0{\hss\sl/\/\hss}}

\def\periodb#1{\setbox0=\hbox{$#1$}#1\hskip-\wd0\hbox to\wd0{-}}
\newcommand{\nablas}{\slasha{\nabla}}

\newcommand{\unit}{\mathbbm{1}}   			
\newcommand{\id}{\mathrm{id}}   			

\newcommand{\CA}{\mathcal{A}}    			

\newcommand{\CC}{\mathcal{C}}
\newcommand{\CCC}{\mathscr{C}}

\newcommand{\CD}{\mathcal{D}}

\newcommand{\CF}{\mathcal{F}}

\newcommand{\CH}{\mathcal{H}}

\newcommand{\CL}{\mathcal{L}}
\newcommand{\CM}{\mathcal{M}}

\newcommand{\CN}{\mathcal{N}}
\newcommand{\CO}{\mathcal{O}}

\newcommand{\CS}{\mathcal{S}}

\newcommand{\frg}{\mathfrak{g}}				

\newcommand{\frX}{\mathfrak{X}}

\newcommand{\Walg}{\mathrm{W}}	
\newcommand{\CEalg}{\mathrm{CE}}

\newcommand{\FR}{\mathbbm{R}}     			
\newcommand{\FC}{\mathbbm{C}}     			
\newcommand{\NN}{\mathbbm{N}}     			
\newcommand{\RZ}{\mathbbm{Z}}     			

\newcommand{\dd}{\mathrm{d}}     			
\newcommand{\dpar}{\partial}     			
\newcommand{\embd}{{\hookrightarrow}}     		
\newcommand*{\longembd}{\ensuremath{\lhook\joinrel\relbar\joinrel\rightarrow}}
\newcommand{\di}{\mathrm{i}}     			
\newcommand{\eps}{{\varepsilon}}			
\newcommand{\epsb}{{\bar{\varepsilon}}}			

\newcommand{\psib}{{\bar{\psi}}}

\newcommand{\ald}{{\dot{\alpha}}}     			
\newcommand{\bed}{{\dot{\beta}}}

\newcommand{\eand}{{\qquad\mbox{and}\qquad}}     		
\newcommand{\ewith}{{\qquad\mbox{with}\qquad}}

\newcommand{\der}[1]{\frac{\dpar}{\dpar #1}}   		
\newcommand{\tr}{\,\mathrm{tr}\,}     			
\newcommand{\ad}{\mathrm{ad}}     			

\newcommand{\au}{\mathfrak{u}}

\newcommand{\aso}{\mathfrak{so}}

\newcommand{\sSp}{\mathsf{Sp}}     			

\newcommand{\sG}{\mathsf{G}}
\newcommand{\sL}{\mathsf{L}}
\newcommand{\sT}{\mathsf{T}}

\newcommand{\sLie}{\mathsf{Lie}}

\newcommand{\sSU}{\mathsf{SU}}

\newcommand{\acton}{\vartriangleright}     			
\renewcommand{\remark}[1]{}     				
\def\tyng(#1){\hbox{\tiny$\yng(#1)$}}			
\def\tyoung(#1){\hbox{\tiny$\young(#1)$}}			

\newcommand{\beq}{\begin{eqnarray}}
\newcommand{\eeq}{\end{eqnarray}}

\newcommand{\sSym}{{\sf Sym}}

\newcommand{\sfs}{{\sf s}}
\newcommand{\sft}{{\sf t}}

\begin{document}
\begin{titlepage}
\begin{flushright}
 DIFA 2015\\
 EMPG--15--17
\end{flushright}
\vskip 2.0cm
\begin{center}
{\LARGE \bf $L_\infty$-Algebra Models and\\[0.5cm] Higher Chern-Simons Theories}
\vskip 1.5cm
{\Large Patricia Ritter$^a$ and Christian S\"amann$^b$}
\setcounter{footnote}{0}
\renewcommand{\thefootnote}{\arabic{thefootnote}}
\vskip 1cm
{\em${}^a$ Dipartimento di Fisica ed Astronomia \\
Universit\`a di Bologna and INFN, Sezione di Bologna\\
Via Irnerio 46, I-40126 Bologna, Italy
}\\[0.5cm]
{\em ${}^b$ Maxwell Institute for Mathematical Sciences\\
Department of Mathematics, Heriot-Watt University\\
Colin Maclaurin Building, Riccarton, Edinburgh EH14 4AS,
U.K.}\\[0.5cm]
{Email: {\ttfamily patricia.ritter@bo.infn.it , C.Saemann@hw.ac.uk}}
\end{center}
\vskip 1.0cm
\begin{center}
{\bf Abstract}
\end{center}
\begin{quote}
We continue our study of zero-dimensional field theories in which the fields take values in a strong homotopy Lie algebra. In a first part, we review in detail how higher Chern-Simons theories arise in the AKSZ-formalism. These theories form a universal starting point for the construction of $L_\infty$-algebra models. We then show how to describe superconformal field theories and how to perform dimensional reductions in this context. In a second part, we demonstrate that Nambu-Poisson and multisymplectic manifolds are closely related via their Heisenberg algebras. As a byproduct of our discussion, we find central Lie $p$-algebra extensions of $\aso(p+2)$. Finally, we study a number of $L_\infty$-algebra models which are physically interesting and which exhibit quantized multisymplectic manifolds as vacuum solutions.
\end{quote}
\end{titlepage}

\tableofcontents

\section{Introduction and results}

Most of our current physical theories, including those describing gravity and string theory, are not truly background independent. That is, their setup includes a given spacetime topology or even a detailed spacetime geometry. It is clear, however, that both topology and geometry of spacetime should ideally emerge from a more fundamental description \cite{Smolin:2005mq}. There are a number of approaches to solve this problem, but the one which is closest to what we have in mind here is that of the IKKT matrix model \cite{Ishibashi:1996xs}, see also \cite{Aoki:1998bq}. This model is equivalent to ten-dimensional super Yang-Mills theory dimensionally reduced to a point and its degrees of freedom are described in a supermultiplet consisting of ten Lie algebra valued matrices and their 16 fermionic superpartners. It has been conjectured that the IKKT model provides a non-perturbative definition of type IIB superstring theory. Interestingly, vacuum solutions of this model and its deformations by background fluxes can be interpreted as noncommutative spaces. Moreover, the physics of small fluctuations around these vacuum solutions corresponds to noncommutative gauge theories. Even the dynamics of the noncommutative spaces themselves can be extracted and related to a theory of noncommutative gravity, see \cite{Yang:2006hj,Steinacker:1003.4134} and references therein.

While this picture is very appealing, there are a number of reasons to consider generalizations in terms of higher Lie algebras or, equivalently, $L_\infty$-algebras: First and  from a mathematician's perspective, it is always a good idea to study deformations and categorifications of mathematical objects to gain a more complete understanding of them. 

From a string theorist's perspective, there are many situations in which ordinary Lie algebras are not sufficient, but have to be replaced by categorified Lie algebras. These range from effective descriptions of M-theory over Kontsevich's deformation quantization to string field theory. For a more comprehensive list with more details and references, we refer to the introduction of \cite{Ritter:2013wpa}. 

From a quantum field theorists perspective, $L_\infty$-algebras appear very naturally in BV-quantization of classical field theories. Moreover, $L_\infty$-algebras come with a canonical equation of motion, the {\em homotopy Maurer-Cartan equation}. This equation is somewhat universal in the sense that many other physical field theories such as Yang-Mills theories and the recently popular M2-brane models of \cite{Bagger:2007jr,Gustavsson:2007vu} can be written as a homotopy Maurer-Cartan equation on some $L_\infty$-algebra, cf.\ e.g.\ \cite{Zeitlin:2007vv,Zeitlin:2007yf,IuliuLazaroiu:2009wz}.

From a gravitational physicist's perspective, there is evidence that
the version of noncommutative gravity obtained from the IKKT model is
too restrictive \cite{Arnlind:2011rz,Arnlind:1312.5454}. This is
perhaps not very surprising, as the noncommutative spaces arising in
the IKKT model are K\"ahler manifolds, and a restriction to those is
clearly insufficient. As suggested in
\cite{Arnlind:2011rz,Arnlind:1312.5454}, one should turn to
Nambu-Poisson manifolds, which have enough overlap with multisymplectic
manifolds as we shall explain in detail. On multisymplectic manifolds,
the Poisson algebra is replaced by a higher Lie algebra of observables
\cite{Baez:2008bu}. Again, we are led to generalizing the IKKT model
to a model involving $L_\infty$-algebras.

It is the latter point that provided most of our motivation to
continue the study of models built from categorified Lie algebras we
began in \cite{Ritter:2013wpa}, considering models employing 2-term
$L_\infty$-algebras. In the present paper, we generalize the
discussion to arbitrary (truncated) $L_\infty$-algebras and
refer to the resulting models as $L_\infty$-algebra models. One of the
most important problems to be solved here is that of selecting a
physically interesting action. Beyond the dimensional reduction of the
bosonic part of the six-dimensional (2,0)-theory, which we will
discuss in section \ref{ssec:2-0-theory}, there is no obvious
candidate. Since other versions of higher Yang-Mills theory are
unknown, we focus on higher versions of Chern-Simons theories. These
theories are conveniently constructed using the language of the
Alexandrov-Kontsevich-Schwarz-Zaboronsky (AKSZ) formalism
\cite{Alexandrov:1995kv}, which we review to the necessary extent. In
this construction, both spacetime and higher gauge algebroid are
regarded as N$Q$-manifolds and a connection is described as a morphism of graded manifolds between them, cf.\
\cite{Bojowald:0406445,Kotov:2007nr,Gruetzmann:2014ica}. The curvature
then measures the failure of this morphism to be a morphism of
N$Q$-manifold. Examples of N$Q$-manifolds include Poisson and
symplectic manifolds as well as Courant algebroids \cite{Roytenberg:0203110}.

We then show how this description generalizes to supersymmetric field theories by employing N$Q$-supermanifolds. In particular, we show how to describe maximally supersymmetric Yang-Mills theory and a recently proposed version of the superconformal field theory in six dimensions \cite{Saemann:2012uq,Saemann:2013pca,Jurco:2014mva}. This is done by encoding the corresponding supermultiplets into a connection on superspace, which is flat along certain subspaces of this superspace. 

In order to follow analogous routes to the construction of the IKKT model, we also study the dimensional reduction of N$Q$-manifolds and its effect on connections encoded in morphisms between them. While dimensional reduction was discussed previously to some degree in \cite{Mayer:2009wf}, see also \cite{Bursztyn:2005vwa} for the case of Courant algebroids, we provide a more complete picture for arbitrary N$Q$-manifolds here. Applying this reduction procedure to the higher Chern-Simons theories obtained previously then yields an interesting set of examples of $L_\infty$-algebra models.

In the second part of this paper, we study solutions of
$L_\infty$-algebra models and connect them to certain higher quantum
geometries. Just as quantizations of certain symplectic manifolds
provide examples of solutions to the IKKT
model, we expect higher quantizations of certain multisymplectic
manifolds to provide classical solutions of the $L_\infty$-algebra
models. While the full quantization of multisymplectic manifolds is still
an open problem, we can discuss solutions using the partial results already available. Note that the quantization map is a homomorphism
of Lie algebras when restricted to the Heisenberg algebra consisting of constant and linear functions. Furthermore, this Heisenberg algebra is sufficient for the description of quantized symplectic manifolds as solutions of the IKKT model. Similarly, we expect that we merely require the Heisenberg $L_\infty$-algebras of multisymplectic manifolds in order to discuss solutions to $L_\infty$-models, which are readily constructed.

As a first step, we review and extend the known connection between
Nambu-Poisson and multisymplectic manifolds. In particular, we point
out that each Nambu-Poisson manifold of degree $p+1$ which is also a $p$-plectic manifold, locally comes with a Heisenberg
$L_\infty$-algebra, which agrees with the Heisenberg
$L_\infty$-algebra that is canonically obtained from the multisymplectic structure. This link is very important in general, because the
mathematically more appealing structures usually arise on the
multisymplectic side, while the physically relevant examples seem to
be based on Nambu-Poisson manifolds. In this context, we also show
that the Heisenberg $L_\infty$-algebra of the sphere $S^n$ is given
by a central extension of $\mathfrak{so}(n+1)$ to an $n-1$-term $L_\infty$-algebra.

After this lengthy setup, we consider various interesting $L_\infty$-algebra models. We start with a Yang-Mills type class of homogeneous $L_\infty$-algebra models, which come with an $L_\infty$-algebra valued set of fields. These comprise the IKKT model and have very similar properties. In particular, we can find higher dimensional analogues of the symplectic manifolds whose Heisenberg algebras form solutions to the IKKT model. 

They differ, however, from inhomogeneous $L_\infty$-algebra models, in which each field takes values in a subspace of an $L_\infty$-algebra with homogeneous grading. Inhomogeneous $L_\infty$-algebra models arise rather naturally in dimensional reductions of higher gauge theories. As an example, we consider the pure gauge part of the $(2,0)$-theory mentioned above. We can readily show that the resulting model is solved by the Heisenberg $L_\infty$-algebra of $\FR^{1,2}\times \FR^3\cong \FR^{1,5}$. This is the categorified version of the fact that the IKKT model is solved by the Heisenberg algebra of $\FR^{1,3}$.

We then show how a very canonical choice of equation of motion arises from dimensionally reducing the higher Chern-Simons action functional: the homotopy Maurer-Cartan equation of the semistrict $L_\infty$-algebra under consideration\footnote{To be precise, it is the homotopy Maurer-Cartan equation of the semistrict Lie $p$-algebra under consideration tensored with the Gra\ss mann algebra $\FR^{1+p}[1]$. Recall that the tensor product of an $L_\infty$-algebra and a differential graded algebra carries a natural $L_\infty$-algebra structure.}. As stated above, the homotopy Maurer-Cartan equation is somewhat universal and we explain in some detail how physically relevant equations of motion are encoded in this equation. 

Because of their universality, it is natural to limit ourselves to homotopy Maurer-Cartan $L_\infty$-algebra models.  Particularly appealing about these models is the fact that the input data consists merely of an $L_\infty$-algebra. We find indeed that a naturally twisted form of these models admits quantizations of $p$-plectic $\FR^{p+1}$ as solutions.

This paper is intended as a starting point for future work, studying the
dynamics of quantized multisymplectic manifolds, along the lines of
\cite{Yang:2006hj,Steinacker:1003.4134}, using $L_\infty$-models.

\section{Metric Lie \texorpdfstring{$p$}{p}-algebroids and symplectic \texorpdfstring{N$Q$}{NQ}-manifolds}

We start with a brief review of some basic concepts and definitions familiar from the AKSZ formalism. In particular, we recall what symplectic N$Q$-manifolds are and how they are related to metric truncated strong homotopy Lie algebras and metric Lie $p$-algebroids.

\subsection{\texorpdfstring{N$Q$}{NQ}-manifolds}\label{ssec:Nq-manifolds}

An {\em N-manifold} is an $\NN$-graded manifold or, and this is where the name stems from, a $\RZ$-graded manifold concentrated in non-negative degrees. The terminology is due to {\v S}evera \cite{Severa:2001aa} and a detailed review can be found in \cite{Roytenberg:0203110}. In general, an $\NN$-graded manifold $\CM$ is a locally ringed space $\CM=(M,\CO_\CM)$, where $M$ is a smooth manifold and $\CO_\CM$ is an $\NN$-graded $\CC^\infty(M)$-sheaf. Alternatively, we can think of an $\NN$-graded manifold $\CM$ being covered by affine charts with coordinates that have each a degree in $\NN$. This implies that the structure sheaf $\CO_\CM$ is filtered: $\CO_\CM=\CO_\CM^0\subset \CO_\CM^1\subset \CO_\CM^2\subset \ldots$, where $\CO_\CM^0=\CC^\infty(M)$ and, more generally, $\CO_\CM^k$ is locally generated by functions of degree $\leq k$. This filtration gives a fibration $\ldots\rightarrow \CM_2\rightarrow \CM_1\rightarrow \CM_0=M$.

A simple, but ubiquitous example of an N-manifold is a parity-shifted linear space, like $\frg[1]$, where $\frg$ is some Lie algebra. The coordinates of $\frg[1]$ all have degree 1 and we say that the N-manifold is concentrated in degree 1. In general, $[n]$ will denote a shift of the grading of some linear space by $n$.

Another class of examples of $N$-manifolds is given by ordinary supermanifolds, which are N-manifolds with grading
concentrated in degrees $0$ and $1$. The canonical example here is
the parity shifted tangent bundle $T[1]M=\Pi TM$ of some manifold $M$,
in which the base manifold is of degree $0$ and the fibers are of
degree 1.  

A more sophisticated example is $T^*[2]T[1]M$, where the
cotangent functor $T^*$ creates fibers with the opposite grading to
those of the base manifold which are then shifted in
degree. Therefore, $T^*[2]T[1]M$ is an N-manifold concentrated in
degrees $0$, $1$ and $2$. This example appears in the interpretation of exact Courant algebroids as N-manifolds \cite{Roytenberg:0203110}.

Recall that a {\em morphism of graded manifolds} $\phi:\CM=(M,\CO_\CM)\rightarrow \CN=(N,\CO_\CN)$ is a degree preserving morphism of ringed spaces. Explicitly, we have a smooth morphism $\phi_0:M\rightarrow N$ together with a morphism $\phi^*:\CO_\CN\rightarrow \CO_\CM$. The latter is fixed by its image on the coordinates generating $\CO_\CN$ locally. Moreover, its degree 0 part is determined by $\phi_0$. This picture generalizes the usual morphism between manifolds: for each coordinate $Z$ on $\CN$, we have a functional dependence $\phi^*(Z)$ on the coordinates of $\CM$.

An {\em N$Q$-manifold} \cite{Severa:2001aa} or {\em differential $\NN$-graded manifold} is now an $\NN$-graded manifold endowed with a homological vector field $Q$. That is, $Q$ is a vector field of degree one which is nilquadratic, $Q^2=0$. {\em N$Q$-morphisms} between two N$Q$-manifolds $(\CM,Q)$ and $(\CM',Q')$ are morphisms of graded manifolds $\varphi:\CM\rightarrow \CM'$ such that $\varphi^*\circ Q'=Q\circ \varphi^*$.

The algebra of functions on an N$Q$-manifold together with the differential $Q$ forms a {\em differential graded algebra}, or a {\em dga} for short. We shall refer to maps between two such differential graded algebras which respect the grading and the differential as {\em dga-morphisms}. Note that morphisms between N$Q$-manifolds are in one-to-one correspondence with dga-morphisms of the corresponding differential graded algebra of functions. For more details on such morphisms, see e.g.\ \cite{Ritter:2015ffa}.

The standard example of an N$Q$-manifold is again given by the shifted tangent bundle $T[1]M$ of some manifold $M$. The algebra of functions on $T[1]M$ and the homological vector field $Q$ are identified with the differential forms $\Omega^\bullet(M)$ on $M$ and the de Rham differential, respectively. The latter is indeed an endomorphism on the algebra of functions on $T[1]M$. 

Note that enhancing the shifted linear space $\frg[1]$ to an
N$Q$-manifold yields a Lie algebra structure on $\frg$. The algebra of
functions on $\frg[1]$ is the graded commutative algebra
$\sSym(\frg[1]^*)$: that is the graded symmetric tensor algebra, freely generated by some coordinates $\xi^\alpha$, $\alpha=1,\ldots,\dim \frg$, of degree $1$ with respect to a basis $\tau_\alpha$ of degree $0$. A vector field $Q$ of degree $1$ is necessarily a differential operator acting on elements of $\sSym(\frg[1]^*)$ as follows:
\begin{equation}
 Q=-\tfrac{1}{2}f^\alpha_{\beta\gamma} \xi^\beta\xi^\gamma\der{\xi^\alpha}~.
\end{equation}
The equation $Q^2=0$ then amounts to the Jacobi identity for the Lie algebra defined by the structure constants $f^\alpha_{\beta\gamma}$. This yields the well-known description of a Lie algebra in terms of its {\em Chevalley-Eilenberg algebra}. The vector field $Q$ is called the {\em Chevalley-Eilenberg differential}.

Finally, we can combine the last two examples. Consider a vector bundle $E\rightarrow M$ over some manifold $M$. Then an N$Q$-structure on $E[1]$ gives rise to a {\em Lie algebroid} \cite{16298602}. Introducing some local coordinates $x^a$ on $M$ and $\xi^\alpha$ in the fibers of $E[1]$, we can write
\begin{equation}
 Q=-m^a_\alpha(x)\xi^\alpha \der{x^a}-\tfrac{1}{2}f^\alpha_{\beta\gamma}(x) \xi^\beta\xi^\gamma\der{\xi^\alpha}~.
\end{equation}
The tensor $f^\alpha_{\beta\gamma}(x)$ defines a Lie algebra structure on the sections of $E$, while the tensor $m^a_\alpha(x)$ encodes the anchor map $\rho:E\rightarrow TM$. The Leibniz rule and the fact that $\rho$ is a morphism of Lie algebras follows from $Q^2=0$.

When discussing supersymmetric field theories, we will have to deal with N$Q$-super\-manifolds. In this case, there is a bigrading of the corresponding algebra of functions. The only thing to keep in mind here is that whenever the order of two odd objects is interchanged, a minus sign has to be inserted. Here, an object is to be regarded as odd, if its total degree, which is the sum of the two degrees of the bigrading, is odd.

\subsection{P-manifolds and symplectic \texorpdfstring{N$Q$}{NQ}-manifolds}

In the context of BV quantization \cite{Schwarz:1992gs,Alexandrov:1995kv}, one often encounters P-manifolds, where P stands for an odd Poisson bracket, which is also called {\em antibracket}. The canonical example here is the supermanifold $T^*[1]\FR^n$ with local coordinates $Z^A=(x^a,p_a)$ parameterizing the base and the fibers, respectively, equipped with the graded Poisson bracket
\begin{equation}\label{eq:ex:Poisson-1}
 \{f,g\}:=f\overleftarrow{\der{x^a}}\overrightarrow{\der{p_a}}g-f\overleftarrow{\der{p_a}}\overrightarrow{\der{x^a}}g~,
\end{equation}
where the arrows indicate derivatives acting from the left and the right. Note that functions on $T^*[1]\FR^n$ can be identified with multivector fields, on which the Poisson bracket \eqref{eq:ex:Poisson-1} acts as the Schouten bracket.

A {\em P$Q$-manifold} \cite{Schwarz:1992gs} is an N$Q$-manifold $\CM$
concentrated in degrees $0$ and $1$, in which the odd Poisson bracket
originates from an odd, $Q$-invariant, non-degenerate 2-form. Given
some coordinates $Z^A$ on $\CM$, $\omega=\tfrac12\dd Z^A\wedge \omega_{AB}\dd Z^B$ is then an odd symplectic form with
\begin{equation}
 \CL_Q\omega=0\eand \{f,g\}=f~\overleftarrow{\frac{\dpar}{\partial Z^A}}\omega^{AB}\overrightarrow{\frac{\dpar}{\partial Z^B}}~g~,
\end{equation}
where $\omega^{AB}$ is the inverse of the graded matrix $\omega_{AB}$.

For example, the Poisson bracket \eqref{eq:ex:Poisson-1} originates from the odd symplectic form $\omega=\dd x^a\wedge \dd p_a$. If we endow $T^*[1]\FR^n$ with the homological vector field $Q=\pi^{ab}p_a\der{x^b}$ with $\pi^{ab}=-\pi^{ba}$ and $Q^2=0$, then $\omega$ is $Q$-invariant:
\begin{equation}
 \CL_Q\omega=\dd \iota_{Q}\omega=\pi^{ab}\dd p_a\wedge \dd p_b=0~.
\end{equation}

There is now a generalized Darboux theorem for P$Q$-manifolds \cite{Schwarz:1992nx} which states that on any P$Q$-manifold $\CM$, we can write $\omega=\dd x^a\wedge \dd p_a$ in some local coordinates $(x^a,p_a)$, $a=1,\ldots,d$.

In this paper, we shall require symplectic forms of general
degree\footnote{By degree of a symplectic form, we shall always mean
  the $\NN$-grading, as the form degree is fixed.}. We thus define a
{\em symplectic N$Q$-manifold of degree $n$} as an N$Q$-manifold
endowed with a symplectic form $\omega$ of degree $n$ satisfying
$\CL_Q\omega=0$. Such structures were called {\em
  $\Sigma_n$-manifolds} in \cite{Severa:2001aa}. A relevant example of
a symplectic N$Q$-manifold of degree 2 is $T[1]M\times_M T^*[2]M$ with
local coordinates $(x^a,\xi^a,p_a)$ and symplectic form $\omega=\dd
x^a\wedge \dd p_a+\tfrac12 g_{ab}\dd \xi^a\wedge \dd \xi^b$, where $g$ is some
invertible matrix.

Let now $\CM$ be a symplectic N$Q$-manifolds of degree $n$ with local coordinates $Z^A=(x^a,\xi^\alpha,\ldots)$ and symplectic form $\omega=\tfrac12\dd Z^A\wedge \omega_{AB}\dd Z^B$. We can associate to any function $f\in \CC^\infty(M)$ a corresponding Hamiltonian vector field $X_f$ via
\begin{equation}
  \label{eq:5}
  X_f\cdot g:=\{g,\, f\}~~~\mbox{or}~~~
  \iota_{X_f}\omega=\dd f~,
\end{equation}
where $\iota_X$ and $\dd$ denote contraction along $X$ and the exterior derivative on $\CM$, respectively. Inversely, we can refer to
$f$ as the \textit{Hamiltonian} of $X_f$. For a function $f$, the parity
of $X_f$ is the opposite of that of $f$ for odd $n$, and
the same for even $n$. Clearly, Hamiltonian vector fields generate symplectomorphisms: $\mathcal{L}_{X_f}\omega=\dd^2 f=0$. 

Note that the homological vector field $Q$ with $Q^2=0$ and $\CL_Q\omega=0$ is the Hamiltonian vector field of a function $\CS$ satisfying $\{\CS,\CS\}=0$. We can compute this function using the 
Euler vector field, which reads as
\begin{equation}\label{eq:Euler_vector_field}
 \eps=\sum_A \deg(Z^A)Z^A\der{Z^A}~,
\end{equation}
expressed locally in coordinates $Z^A$. A
closed, homogeneous form $\lambda$ of degree $m$, i.e.\ such that
$\mathcal{L}_\varepsilon\lambda=m \lambda$ and $\dd \lambda=0$, is exact, that
is $\lambda=\dd\varphi$, with $\varphi=\tfrac{1}{m}\iota_\varepsilon
\lambda$. In particular, one can check that $\mathcal{L}_\varepsilon
\omega=n\omega$, for the symplectic structure of degree $n$, so
that the symplectic potential is
$\alpha=\tfrac{1}{n}\iota_\varepsilon\omega$, and
$\dd\alpha=\omega$. Similarly one can compute that
$\mathcal{L}_\varepsilon\iota_Q\omega=(n+1)\iota_Q\omega$. But
$\iota_Q\omega$ is, by definition, the exterior derivative of the
Hamiltonian $\CS$ of $Q$: $\iota_Q\omega=\dd\CS$.
Explicitly, we therefore have (see e.g.\ \cite{Fiorenza:2011jr})
\begin{equation}
 \CS=\frac{1}{1+n}\iota_\eps\iota_Q\omega~.
\end{equation}
Conversely, any function $\CS$ with $\{\CS,\CS\}=0$ has a homological Hamiltonian vector field.

Note that we have the following tower of symplectic N$Q$-manifolds of degree $n$. In the case $n=0$, non-degeneracy of the symplectic structure implies that the N$Q$-manifold is concentrated in degree 0 and we thus obtain an ordinary symplectic manifold. As stated above, for $n=1$, the N$Q$-manifold is necessarily of the form $T^*[1]M$ and the Hamiltonian is a bivector field on $M$. We thus obtain a Poisson manifold. Continuing along these lines, we find that $n=2$ yields a Courant (2-)algebroid \cite{Severa:2001aa,Roytenberg:0203110}, as we shall review later.

As a final example, let $\CM = T^*[1]\CN$ be the shifted cotangent bundle of the manifold $\CN=X\times \mathfrak{g}[1]$, where $X$ is a $\sG$-manifold for some Lie group $\sG$ with Lie algebra $\mathfrak{g}$. The coordinates on $\CM$ are $Z^A=(x^a,p_a,\,\xi^\alpha,\,\zeta_\alpha)$ with degree $(0,\,1,\,1,\,0)$
respectively. The standard P-structure on $T^*[1]N$ is then
\begin{equation}
  \label{eq:7}
  \omega=\tfrac12\dd Z^A\wedge \dd Z_A=\dd x^a\wedge \dd p_a+\dd \xi^\alpha\wedge\dd \zeta_\alpha~,
\end{equation}
while a nilpotent $Q$, of degree 1, preserving $\omega$ is necessarily of the form
\begin{equation}
  \label{eq:11}
Q=-T^a_\alpha \xi^\alpha\frac{\partial}{\partial
  x^a}-\tfrac{1}{2}f^\alpha_{\beta\gamma}\xi^\beta \xi^\gamma\frac{\partial}{\partial
  \xi^\alpha}+\frac{\partial T^a_\alpha}{\partial
  x^b}p_a\xi^\alpha\frac{\partial}{\partial p_b} + \left( T^a_\beta p_a
  + 2 
  f^\alpha_{\beta\gamma}\zeta_\alpha\xi^\gamma\right)\frac{\partial}{\partial \zeta_\beta}~.
  \end{equation}
Here, the $f^\alpha_{\beta\gamma}$ are the structure constants of the Lie algebra
$\mathfrak{g}$ in a basis $(\tau_\alpha)$, where a general element of
$\mathfrak{g}[1]$ is of the form $\xi^\alpha \tau_\alpha$. The anchor map of a Lie algebroid structure over $X\times \frg^*$ is encoded in the first and last summand of $Q$. We arrive at a symplectic N$Q$-manifold of degree 1, which is of the form $T^*[1]M\cong T^*[1]\CN$ for $M=X\times \frg[0]^*$.

\subsection{\texorpdfstring{$L_\infty$}{L-infinity}-algebras and \texorpdfstring{$L_\infty$}{L-infinity}-algebroids}

To describe higher gauge theory leading to $L_\infty$-algebra models
later, we have to generalize our discussion from Lie algebroids to (symplectic) Lie
$p$-algebroids. The latter are iterative categorifications of the
notion of a Lie algebroid and they are most easily described using the analogue of the Chevalley-Eilenberg description of Lie
algebras.

As a first step, let us consider the Chevalley-Eilenberg description
of an {\em $L_\infty$-algebra}. Here, we have a differential graded
algebra $\sSym(\frg[1]^*)$, where
$\frg$ is a linear N-manifold with global coordinates $Z^A$ of
degree $|A|\in \NN$ in some basis $\tau_A$ of degree 0. The homological vector field $Q$ on the N$Q$-manifold $\frg[1]$ is of the general form
\begin{equation}
 Q=-m_B^A  Z^B \der{Z^A}-\frac{1}{2}m_{BC}^A Z^B Z^C \der{Z^A}+\ldots=\sum_{k=1}^\infty \frac{(-1)^{\sigma_k}}{k!}m^A_{B_1\ldots B_k} Z^{B_1}\ldots Z^{B_k}\der{Z^A} ~,
\end{equation}
where
\begin{equation}\label{eq:sigma_k}
 \sigma_k:=\binom{k+1}{2}=\frac{k(k+1)}{2}~.
\end{equation}
The minus signs are inserted for later convenience. The fact that $Q$ is of degree one restricts the type of non-vanishing generalized structure constants $m^A_{B_1\ldots B_k}$. They define a set of brackets on elements of $\frg$, which leads to the dual picture of an $L_\infty$-algebra. We now also introduce the graded basis $\hat \tau_A$ on $\frg$: here all coordinates are of degree 0, while the $\hat\tau_A$ absorb the grading of $Z^A$ pre-shift. To match common conventions in the literature, we choose $\hat\tau_A$ such that $|\hat\tau_A|=|A|-1$. We then  get the brackets
\begin{equation}\label{NPQ-mfd and n-algebras, brackets}
 \mu_k(\hat \tau_{B_1},\ldots,\hat \tau_{B_k}):=m^A_{B_1\ldots B_k}\hat \tau_A~,
\end{equation}
where each bracket $\mu_k$ itself has degree $k-2$. The condition $Q^2=0$ directly translates to homotopy Jacobi identities, and together with the brackets, $\frg$ forms an $L_\infty$- or strong homotopy Lie algebra \cite{Lada:1992wc}. At lowest order, we have
\begin{equation}
 \mu_1(\mu_1(\hat \tau_A))=0~,~~~\mu_1(\mu_2(\hat \tau_A,\hat \tau_B))=\mu_2(\mu_1(\hat \tau_A),\hat \tau_B)\pm \mu_2(\hat \tau_A,\mu_1(\hat \tau_B))~,~~~\ldots
\end{equation}
The $\pm$-sign in the last term has to be chosen according to the grading of $\hat \tau_A$. 

If the N$Q$-manifold is concentrated in degrees $1$ to $p$, the resulting $L_\infty$-algebra is truncated and it is categorically equivalent to a (semistrict) Lie $p$-algebra. 

Morphisms of $L_\infty$-algebras and Lie $p$-algebras are now simply morphisms of the underlying N$Q$-manifolds, see also \cite{Ritter:2015ffa} and references therein. 

The most general situation of an {\em $L_\infty$-algebroid} is
obtained by considering an arbitrary N$Q$-manifold $\CM$. Contrary to
the last example, the degree $0$-component $M$ of $\CM$ will be
non-trivial, and we will refer to this component as the body $\CM_0$
of $\CM$. At a point $x\in \CM_0$, we can introduce a system of
coordinates $Z^A=(x^a,\xi^\alpha,\ldots)$ and express the homological
vector field $Q$ in the form\footnote{For simplicity, we assume here
  that $x$ is a stationary point, so that the sum starts at $k=1$
  instead of $k=0$.}
\begin{equation}
\begin{aligned}
 Q&=-m_B^A(x) Z^B \der{Z^A}-\frac{1}{2}m_{BC}^A(x) Z^B Z^C \der{Z^A}+\ldots\\
 &=-\sum_{k=1}^\infty \frac{(-1)^{\sigma_k}}{k!}m^A_{B_1\ldots B_k}(x) Z^{B_1}\ldots Z^{B_k}\der{Z^A}~.
\end{aligned}
\end{equation}
Note that here, the coefficients $m^A_{B_1\ldots B_k}(x)$ are now sections over $\CM_0$. Again, by concentrating the N$Q$-manifolds in degrees $d$ with $0\leq d\leq p$, we obtain Lie $p$-algebroids. 

\subsection{Metric \texorpdfstring{$L_\infty$}{L-infinity}-algebroids}\label{ssec:metric_L_infty_algebroids}

Adding a symplectic structure to an N$Q$-manifold adds a metric to the corresponding $L_\infty$-algebroid. Since the discussion of a metric will be crucial later on, we shall be rather explicit in explaining this fact.

We first restrict ourselves to $L_\infty$-algebras arising from N$Q$-manifolds concentrated in degrees $d\geq 1$. Recall that a {\em cyclic inner product} on an $L_\infty$-algebra $L$ is a non-degenerate graded symmetric bilinear map $(-,-):L\times L\rightarrow \FR$ such that 
\begin{equation}\label{eq:ax_cyclic_metric}
\begin{aligned}
 (x_1,x_2)&=(-1)^{\tilde x_1\,\tilde x_2}(x_2,x_1)~,\\
 (\mu_k(x_1,\ldots,x_k),x_0)&=(-1)^{k+\tilde x_0(\tilde x_1+\cdots+\tilde x_k)}\big(\mu_k(x_0,x_1,\ldots,x_{k-1}),x_k\big)~.
\end{aligned}
\end{equation}
For the original references and more details on cyclic inner products, see \cite{Zwiebach:1992ie,Kontsevich:1992aa,Igusa:2003yg,0821843621}.

Take an N$Q$-manifold $\CM$, without body, which is endowed with a symplectic form $\omega$ satisfying $\CL_Q\omega=0$. This symplectic form gives rise to a cyclic inner product on the corresponding $L_\infty$-algebra. In particular, if $n$ is the degree of $\omega$, we obtain a Lie $p$-algebra with $p=n-1$. The latter relation between $p$ and $n$ will hold throughout rest of the paper.

Instead of considering the general case, let us illustrate this statement using the case of a Lie 2-algebra. Take a symplectic N$Q$-manifold $\CM=(\CM_2 \rightarrow \CM_1\rightarrow *)$, where $\CM_1=V[1]$ and $(\CM_2\rightarrow \CM_1)=T^*[3]V[1]$. Let us introduce coordinates $(\xi^\alpha,p_\alpha)$ on $T^*[3]V[1]$. Then $Q$ reads as
\begin{equation}\label{eq:Q_ex_1}
 Q=-e^{\alpha\beta} p_\beta \der{\xi^\alpha}-\frac12 f^\alpha_{\beta\gamma}\xi^\beta\xi^\gamma\der{\xi^\alpha}- g^\gamma_{\beta \alpha}\xi^\beta p_\gamma\der{p_\alpha}+\frac{1}{3!}h_{\alpha\beta\gamma\delta}\xi^\alpha\xi^\beta\xi^\gamma \der{p_\delta}~,
\end{equation}
while the symplectic form is given by 
\begin{equation}
 \omega=\dd \xi^\alpha\wedge \dd p_\alpha~,~~~\CL_Q\omega=0~.
\end{equation}
For the graded basis vectors $\hat \tau_A=\{\hat \tau_\alpha,\hat
\tau^\beta\}$ of $\CM[-1]=V\oplus V^*[1]$, which generate the $L_\infty$-algebra, we have the corresponding higher brackets
\begin{equation}
 \mu_1(\hat \tau^\alpha)=e^{\alpha\beta} \hat \tau_\beta~,~~~\mu_2(\hat \tau_\alpha,\hat \tau_\beta)=f^\gamma_{\alpha\beta}\hat \tau_\gamma~,~~~\mu_2(\hat \tau_\alpha,\hat \tau^\beta)=g^\beta_{\alpha\gamma}\hat \tau^\gamma~,~~~\mu_3(\hat \tau_\alpha,\hat \tau_\beta,\hat \tau_\gamma)=h_{\alpha\beta\gamma\delta}\hat \tau^\delta~.
\end{equation}
The fact that $Q^2=0$ implies the following homotopy relations on the higher brackets:
\begin{subequations}\label{eq:homotopy_relations}
\begin{equation}
\begin{aligned}
 \mu_1(\mu_2(\hat \tau_\alpha,\hat \tau^\beta))&=\mu_2(\hat \tau_\alpha,\mu_1(\hat \tau^\beta))~,~~~\mu_2(\mu_1(\hat \tau^\alpha),\hat \tau^\beta)=\mu_2(\hat \tau^\alpha,\mu_1(\hat \tau^\beta))~,\\
 \mu_3(\hat \tau^\alpha,\hat \tau^\beta,\hat \tau^\gamma)&=\mu_3(\hat \tau^\alpha,\hat \tau^\beta,\hat \tau_\gamma)=\mu_3(\hat \tau^\alpha,\hat \tau_\beta,\hat \tau_\gamma)=0~,\\
 \mu_1(\mu_3(\hat \tau_\alpha,\hat \tau_\beta,\hat \tau_\gamma))&=-\mu_2(\mu_2(\hat \tau_\alpha,\hat \tau_\beta),\hat \tau_\gamma)-\mu_2(\mu_2(\hat \tau_\gamma,\hat \tau_\alpha),\hat \tau_\beta)-\mu_2(\mu_2(\hat \tau_\beta,\hat \tau_\gamma),\hat \tau_\alpha)~,\\
 \mu_3(\mu_1(\hat \tau^\alpha),\hat \tau_\beta,\hat \tau_\gamma)&=-\mu_2(\mu_2(\hat \tau_\beta,\hat \tau_\gamma),\hat \tau^\alpha)-\mu_2(\mu_2(\hat \tau^\alpha,\hat \tau_\beta),\hat \tau_\gamma)-\mu_2(\mu_2(\hat \tau_\gamma,\hat \tau^\alpha),\hat \tau_\beta) 
\end{aligned}
\end{equation}
and
\begin{equation}
\begin{aligned}
 \mu_2(\mu_3(\hat \tau_\alpha,&\hat \tau_\beta,\hat \tau_\gamma),\hat \tau_\delta)-\mu_2(\mu_3(\hat \tau_\delta,\hat \tau_\alpha,\hat \tau_\beta),\hat \tau_\gamma)+\mu_2(\mu_3(\hat \tau_\gamma,\hat \tau_\delta,\hat \tau_\alpha),\hat \tau_\beta)\\
 & -\mu_2(\mu_3(\hat \tau_\beta,\hat \tau_\gamma,\hat \tau_\delta),\hat \tau_\alpha)=\\
 &\mu_3(\mu_2(\hat \tau_\alpha,\hat \tau_\beta),\hat \tau_\gamma,\hat \tau_\delta)-\mu_3(\mu_2(\hat \tau_\beta,\hat \tau_\gamma),\hat \tau_\delta,\hat \tau_\alpha)+\mu_3(\mu_2(\hat \tau_\gamma,\hat \tau_\delta),\hat \tau_\alpha,\hat \tau_\beta)\\
 &-\mu_3(\mu_2(\hat \tau_\delta,\hat \tau_\alpha),\hat \tau_\beta,\hat \tau_\gamma)
 -\mu_3(\mu_2(\hat \tau_\alpha,\hat \tau_\gamma),\hat \tau_\beta,\hat \tau_\delta)-\mu_3(\mu_2(\hat \tau_\beta,\hat \tau_\delta),\hat \tau_\alpha,\hat \tau_\gamma)~.
\end{aligned}
\end{equation}
\end{subequations}
The inner product on the graded basis vectors $\hat \tau_A$ reads as
$(\hat \tau_\alpha,\hat \tau^\beta)=(\hat \tau^\beta,\hat
\tau_\alpha)=\delta_\alpha^\beta$. Because of $\CL_Q\omega=0$, this inner product is indeed cyclic. We have
\begin{equation}
 \begin{aligned}
  0&=\CL_Q(\dd \xi^\alpha\wedge \dd p_\alpha)=\big(\dd (Q(\xi^\alpha))\wedge \dd p_\alpha+\dd \xi^\alpha\wedge \dd (Q(p_\alpha))\big)\\
  &=-\dd (p_\beta e^{\alpha\beta}+\tfrac12 f^\alpha_{\beta\gamma}\xi^\beta\xi^\gamma)\wedge \dd p_\alpha-\dd \xi^\alpha\wedge\dd (g^{\gamma}_{\beta \alpha}\xi^\beta p_\gamma -\tfrac{1}{3!}h_{\beta\gamma\delta\alpha}\xi^\beta\xi^\gamma\xi^\delta)~,
 \end{aligned}
\end{equation}
which implies that
\begin{equation}
\begin{aligned}
  (\mu_1(\hat \tau^\alpha),\hat \tau^\beta)=e^{\alpha\beta}&=e^{\beta\alpha}=(\mu_1(\hat \tau^\beta),\hat \tau^\alpha)~,\\
  (\mu_2(\hat \tau_\alpha,\hat \tau_\beta),\hat \tau^\gamma)=f^\gamma_{\alpha\beta}&=-g^\gamma_{\alpha\beta}=(\mu_2(\hat \tau^\gamma,\hat \tau_\alpha),\hat \tau_\beta)\\
  &=g^\gamma_{\beta\alpha}=(\mu_2(\hat \tau_\beta,\hat \tau^\gamma),\hat \tau_\alpha)~,\\
  (\mu_3(\hat \tau_\alpha,\hat \tau_\beta,\hat \tau_\gamma),\hat \tau_\delta)=h_{\alpha\beta\gamma\delta}&=-h_{\delta\alpha\beta\gamma}=-(\mu_3(\hat \tau_\delta,\hat \tau_\alpha,\hat \tau_\beta),\hat \tau_\gamma)~.
\end{aligned}
\end{equation}

The generalization to $L_\infty$-algebroids is readily performed. To illustrate this, we consider the case of symplectic Lie 2-algebroids or Courant algebroids, see also \cite{Roytenberg:0203110}. Here, we have a graded manifold $\CM=E[1]\times_M T^*[2]M=\CM_2\rightarrow \CM_1\rightarrow \CM_0$, where $E$ is some vector bundle over a manifold $M$. We introduce local Darboux coordinates $(x^a,\xi^\alpha,p_a)$ with grading $0$, $1$ and $2$, respectively, in which the symplectic form of degree 2 reads as
\begin{equation}
 \omega=\dd x^a\wedge \dd p_a+\tfrac{1}{2}\dd \xi^\alpha\wedge \omega_{\alpha\beta}\dd \xi^\beta~.
\end{equation}
A homological vector field $Q$ compatible with $\omega$ (i.e.\ a $Q$ such that $\CL_Q\omega=0$) is necessarily of the form
\begin{equation}
 \begin{aligned}
    Q=-m^a_\alpha\xi^\alpha \frac{\partial}{\partial x^a}
  &-\left(\frac12 m^\alpha_{\gamma\beta}\xi^\gamma\xi^\beta +m^a_\alpha p_a \omega^{\beta\alpha}\right)\frac{\dpar}{\dpar\xi^\alpha}+\\
  &+\left(\frac{1}{3!}\frac{\dpar m^\delta_{\beta\gamma}}{\dpar x^a}\omega_{\delta\alpha}\xi^\alpha\xi^\beta\xi^\gamma+
  \frac{\dpar m^b_\alpha}{\dpar x^a}p_b\xi^\alpha\right)\frac{\dpar}{\dpar p_a}~.
 \end{aligned}
\end{equation}
The above data now encodes a Courant algebroid. The relevant vector
bundle $E$ is $\CM_1\rightarrow \CM_0$, i.e.\ the vector bundle $E[1]$
we started from. A section $\xi$ of this bundle is locally given by
functions $\xi^\alpha(x)$ and the anchor map $\varrho:E\rightarrow TM$
is defined via $\varrho(\hat \tau_\alpha)=m_\alpha^a\der{x^a}$. The inner
product between fiber elements $\xi_1=\xi_1^\alpha \tau_\alpha$ and
$\xi_2=\xi_2^\alpha \tau_\alpha$ in some (degree 0) basis $(\tau_\alpha)$ of the fibers is given by $(\xi_1,\xi_2):=\xi_1^\alpha\omega_{\alpha\beta}\xi_2^\beta$. Note that the anchor map also defines a pullback $\CD$ of the exterior derivative $\dd$ on $\CM_0$ via the adjoint map
\begin{equation}
 (\CD f)^\alpha=\tfrac12\omega^{\alpha\beta}m^a_\beta\der{x^a}f~.
\end{equation}
In terms of the graded basis $\hat\tau_\alpha$ in which the coordinates have degree 0, the Courant bracket on sections is $\mu_2(\hat \tau_\alpha,\hat \tau_\beta)=m^\gamma_{\alpha\beta}\hat \tau_\gamma$. Due to $\CL_Q\omega=0$, its associator satisfies
\begin{equation}
 \mu_2(\mu_2(\hat \tau_\alpha,\hat \tau_\beta),\hat \tau_\gamma)+\mu_2(\mu_2(\hat \tau_\beta,\hat \tau_\gamma),\hat \tau_\alpha)+\mu_2(\mu_2(\hat \tau_\gamma,\hat \tau_\alpha),\hat \tau_\beta)+\tfrac12\CD(\mu_2(\hat \tau_{[\alpha},\hat \tau_\beta),\hat \tau_{\gamma]})=0~,
\end{equation}
where the last term is a cyclic sum. The remaining axioms of Courant algebroids read as
\begin{equation}
 \begin{aligned}
 \varrho(\mu_2( \xi_1,\xi_2))&=[\varrho(\xi_1),\varrho(\xi_2)]~,\\
 \mu_2( \xi_1,f \xi_2)&=f\mu_2( \xi_1,\xi_2)+(\varrho(\xi_1)\cdot f)\xi_2-( \xi_1,\xi_2)\CD f~,\\
 \varrho(\xi_1)\cdot( \xi_2,\xi_3) &=\big(\mu_2( \xi_1,\xi_2)+\CD( \xi_1,\xi_2),\xi_3 \big)+ \big( \xi_2 ,\mu_2( \xi_1,\xi_3)+\CD( \xi_1 ,\xi_3) \big)~,\\
 ( \CD f,\CD g)&=0
 \end{aligned}
\end{equation}
for all $\xi_1,\xi_2,\xi_3\in \Gamma(\CM_1\rightarrow \CM_0)$ and $f\in \CC^\infty(\CM_0)$. These are readily verified to hold, too. Note that for $\CM_0$ a point, we obtain a metric Lie algebra.

\subsection{Weil algebras and invariant polynomials}\label{ssec:Weil_algebra}

The Weil algebra $\Walg(\frg)$ of an $L_\infty$-algebroid $\frg$ is the Chevalley-Eilenberg algebra of the tangent $L_\infty$-algebroid $T[1]\frg$, endowed with a canonical Chevalley-Eilenberg differential, cf.\ \cite{Sati:0801.3480}. For example, the Weil algebra of the trivial Lie algebroid consisting of a manifold $M$ is the algebra of functions on the tangent Lie algebroid $T[1]M$, which is given by the de Rham complex $\Walg(M)=\CEalg(TM)=(\Omega^\bullet(M),\dd)$.

Here, we will be interested in the Weil algebra of an
$L_\infty$-algebra $\frg$. As a vector space, the tangent $L_\infty$-algebra $T[1]\frg$ can be identified with $\frg\oplus \frg[1]$ and we obtain the Weil algebra as the graded vector space
\begin{equation}
 \sSym(\frg[1]^*\oplus \frg[2]^*)~.
\end{equation}
Let $\dd_\frg$ be the de Rham differential on $\frg$, inducing a shift isomorphism $\dd_\frg:\frg[1]^*\rightarrow \frg[2]^*$. The canonical Chevalley-Eilenberg differential $\dd_\Walg$ on $T[1]\frg[1]$ acts on the coordinates $\xi^\alpha$ of $\frg[1]$ as follows:
\begin{equation}
 \dd_\Walg \xi^\alpha=\dd_\CEalg \xi^\alpha+\dd_\frg\xi^\alpha\eand
 \dd_\Walg\,\dd_\frg\,\xi^\alpha=-\dd_\frg\,\dd_\CEalg\,\xi^\alpha~.
\end{equation}
Note that for an ordinary Lie algebra, this reproduces the conventional definition of the Weil algebra.

The Weil algebra has trivial cohomology and fits into the sequence
\begin{equation}
 {\rm inv}(\frg) \ \embd\ \Walg(\frg) \ \xrightarrow{\pi_\Walg} \ \CEalg(\frg)~,
\end{equation}
where $\pi_W$ is the obvious projection $\sSym(\frg[1]^*\oplus \frg[2]^*)\rightarrow \sSym(\frg[1]^*)$. The {\em invariant polynomials} ${\rm inv}(\frg)$ are elements of the Weil algebra that sit completely in $\sSym(\frg[2]^*)$ and are closed under $\dd_\Walg$. Therefore, the obvious contraction of an element $p\in {\rm inv}(\frg)$ with an element of $X\in\frg$ vanishes and so does its Lie derivative $\CL_X:=[\dd_\Walg,\iota_X]$, which encodes the coadjoint action of $X$ on elements of $\Walg(\frg)$. This justifies referring to these elements as invariant polynomials.

For the trivial Lie algebroid consisting of a manifold $M$, the invariant polynomials are closed differential forms of minimal degree $1$. For an ordinary Lie algebra $\frg$, we recover the usual invariant polynomials. Recall that these are given by expressions $p(\tau_1,\tau_2,\ldots,\tau_n)$ such that $p([\tau_0,\tau_1],\tau_2,\ldots,\tau_n)+\ldots+p(\tau_1,\tau_2,\ldots,[\tau_0,\tau_n])=0$ for elements $\tau_0,\ldots,\tau_n\in \frg$.

There is now a {\em transgression} between a Chevalley-Eilenberg cocycle and an invariant polynomial going through a {\em Chern-Simons element} of the Weil algebra: Given a cocycle $\kappa\in {\rm CE}(\frg)$ with $\dd_{\rm CE} \kappa=0$ as well as an invariant polynomial $\hat\omega\in {\rm inv}(\frg)$, then $\chi$ is called a corresponding Chern-Simons element if $\dd_W \chi=\hat\omega$ and $\chi|_{\rm CE}=\kappa$. Note that this type of transgression is closely related to the transgression of forms on the typical fiber of a fiber bundle to forms on its base. 

As a straightforward example, consider the symplectic N$Q$-manifold
$(\frg[1],\omega)$ concentrated in degree 1, where $\omega=\tfrac12\dd
\xi^\alpha\wedge \omega_{\alpha\beta}\dd \xi^\beta$, in some coordinates $\xi^\alpha$ on
$\frg[1]$, and $Q=\dd_\CEalg=-\tfrac12 f_{\alpha\beta}^\gamma\xi^\alpha\xi^\beta \tfrac{\partial}{\partial\xi^\gamma}$. On $W(\frg[1])=\CEalg(T[1]\frg[1])$ with generators $\dd_\frg \xi^\alpha$ in the fibers and $\xi^\alpha$ on the base, the invariant polynomial corresponding to $\omega$ is $\hat \omega=\tfrac12\dd_\frg \xi^\alpha\wedge \omega_{\alpha\beta}\dd_\frg\xi^\beta$. Note that the Hamiltonian $\CS=-\tfrac{1}{3!}\omega_{\alpha\delta}f^\delta_{\beta\gamma}\xi^\alpha\xi^\beta\xi^\gamma$ of the homological vector field $Q$ is indeed a Chevalley-Eilenberg cocycle, since $Q\CS=\{\CS,\CS\}=0$ by definition. The invariant polynomial corresponding to $\omega$ can now be obtained as a transgression of the Chevalley-Eilenberg cocycle $\CS$ via the Chern-Simons element
\begin{equation}
 \chi=\tfrac12\left(\omega_{\alpha\beta}\xi^\alpha \dd_\frg\xi^\beta-\tfrac{1}{3!}\omega_{\alpha\delta}f^\delta_{\beta\gamma}\xi^\alpha\xi^\beta\xi^\gamma\right)~,
\end{equation}
because we have $\dd_{\rm W}\chi=\hat \omega$ and evidently $\chi|_{\rm CE(\frg)}=\frac{1}{n}\CS$.

We can generalize this to arbitrary symplectic N$Q$-manifolds $\CM$ with local Darboux coordinates $Z^A$ and symplectic form $\omega=\tfrac12\dd Z^A\wedge \omega_{AB}\dd Z^B$ of degree $n$. As shown in \cite{Fiorenza:2011jr}, the cocycle $\frac{1}{n}\CS=\frac{1}{n(n+1)}\iota_\eps\iota_Q\omega$ can be transgressed to the invariant polynomial corresponding to the symplectic form, via the Chern-Simons element $\chi=\frac{1}{n}(\iota_\eps\omega+\CS)$, where $\eps$ is the Euler vector field \eqref{eq:Euler_vector_field}. We will present the details for $n=3$ in section \ref{sec:actions-aksz-sigma}.

\section{Higher gauge theory with symplectic \texorpdfstring{N$Q$}{NQ}-manifolds}

Higher gauge theory \cite{Baez:2002jn,Baez:2004in,Baez:0511710,Baez:2010ya} describes the parallel transport of extended objects, just as ordinary gauge theory describes the parallel transport of point-like objects. In this section, we concisely review the formalism of \cite{Bojowald:0406445,Kotov:2007nr,Gruetzmann:2014ica}, see also \cite{Roytenberg:2006qz}. This formalism describes higher gauge theories in terms of morphisms between N$Q$-manifolds, generalizing the ideas of \cite{Atiyah:1957} and \cite{Alexandrov:1995kv}. We also make contact with the discussion in \cite{Sati:0801.3480,Fiorenza:2011jr}. 

\subsection{Motivating example: Ordinary gauge theory}

Let us start by reformulating the local kinematical data of ordinary
gauge theory in the language of N$Q$-manifolds. A connection on a topologically trivial principal
fiber bundle $P$ with structure group $\sG$ over a manifold $\Sigma$ can be
encoded in a 1-form $A$ taking values in the Lie algebra
$\frg=\sLie(\sG)$. Its curvature is $F=\dd A+\tfrac12 [A,A]$ and gauge
transformations are encoded in $\sG$-valued functions $g$ which act on $A$ according to $A\mapsto \tilde A=g^{-1}Ag+g^{-1}\dd g$. At infinitesimal level, we have $A\mapsto A+\delta A$, $\delta A=\dd \eps+[A,\eps]$, where $\eps$ is a $\frg$-valued function.

Note that the shifted tangent bundle $T[1]\Sigma$ forms an N$Q$-manifold with $Q=\dd_\Sigma$ being the de Rham differential on $\Sigma$. Functions on $T[1]\Sigma$ are differential forms, and $\dd_\Sigma$ is therefore indeed a vector field on $T[1]\Sigma$ of degree one. The gauge algebra $\frg$ is also regarded as an N$Q$-manifold $\frg[1]$ with $Q=\dd_{\rm CE}$ being the Chevalley-Eilenberg differential. Recall that on some coordinates $\xi^\alpha$ on $\frg[1]$ with respect to a basis $\tau_\alpha$, $Q$ acts according to 
\begin{equation}
 Q\xi^\alpha= \dd_\CEalg \xi^\alpha=-\tfrac12 f^\alpha_{\beta\gamma}\xi^\beta\xi^\gamma~,
\end{equation}
where $f^\alpha_{\beta\gamma}$ are the structure constants of $\frg$, cf.\ section \ref{ssec:Nq-manifolds}.

The Lie algebra valued 1-form $A$ is then encoded in a morphism of graded manifolds $a:T[1]\Sigma\rightarrow \frg[1]$. Since $\frg[1]$ is concentrated in degree 1, this morphism is fully characterized by a map $a^*:\xi^\alpha\rightarrow A^\alpha$, where $A=A^\alpha\tau_\alpha$ is a $\frg$-valued 1-form on $\Sigma$. The curvature $F=F^\alpha\tau_\alpha$ of $A$ corresponds now to the failure of $a$ to be a morphism of N$Q$-manifolds:
\begin{equation}
\begin{aligned}
 F^\alpha&=(\dd_\Sigma\circ a^*-a^*\circ Q)(\xi^\alpha)\\
 &=\dd_\Sigma a^*(\xi^\alpha)-a^*(-\tfrac12 f^\alpha_{\beta\gamma}\xi^\beta\wedge \xi^\gamma)=\dd_\Sigma A^\alpha+\tfrac12 f^\alpha_{\beta\gamma}A^\beta\wedge A^\gamma~.
\end{aligned}
\end{equation}

Since gauge potentials are described by morphisms of graded manifolds, it is only natural to describe gauge transformations in terms of flat homotopies between these \cite{Bojowald:0406445,Fiorenza:2010mh}. That is, given two gauge equivalent gauge potentials in terms of morphisms $a,\tilde a$ of graded manifolds $T[1]\Sigma\rightarrow \frg[1]$, we lift these to a morphism of graded manifolds,
\begin{equation}
 \hat a: T[1](\Sigma\times[0,1])\rightarrow \frg[1]~,
\end{equation}
such that the restrictions to the endpoints of $[0,1]$ yield $a$ and $\tilde a$. More explicitly, introducing coordinates $x^\mu$ and $r$ on $\Sigma$ and $[0,1]$, respectively, we demand that $\hat{a}(x,r)|_{r=0}=a(x)$ and $\hat{a}(x,r)|_{r=1}=\tilde{a}(x)$. The morphism $\hat a$ is fully characterized by a map $\hat a:\xi^\alpha \rightarrow \hat A^\alpha$, where $\hat A^\alpha\tau_\alpha$ is now a $\frg$-valued 1-form $\Sigma\times [0,1]$ with components
\begin{equation}
 \hat A^\alpha=\hat A^\alpha_\mu \dd x^\mu+\hat A^\alpha_r \dd r~.
\end{equation}
Its curvature which is calculated with respect to the augmented homological vector field $\hat Q=\dd_\Sigma+\dd_{[0,1]}$ has components
\begin{equation}
\hat F=\tfrac12\hat F_{\mu\nu}\dd x^\mu\wedge \dd x^\nu+\underbrace{\left(\der{x^\mu}\hat A_r(x,r)+\mu_2(\hat A_\mu(x,r),\hat A_r(x,r))-\der{r}\hat A_\mu(x,r)\right)\dd x^\mu\wedge \dd r}_{\hat{F}_\perp}~,
\end{equation}
and for the homotopy to be flat, $\hat F_\perp$ has to vanish. This implies that 
\begin{equation}\label{eq:inft_gauge}
 \der{r}\hat A_\mu(x,r)=\der{x^\mu}\hat A_r(x,r)+\mu_2(\hat A_\mu(x,r),\hat A_r(x,r))~,
\end{equation}
and we recover the usual infinitesimal gauge transformations with gauge parameter $A_r(x,0)$. The differential equation \eqref{eq:inft_gauge} can now be integrated to obtain finite gauge transformations.

In the global picture, we would consider the Atiyah Lie algebroid associated to the principal fiber bundle $P$ over $\Sigma$ \cite{Atiyah:1957}. This algebroid $T[1]P/G$ sits in the exact sequence of vector bundles
\begin{equation}
 0\rightarrow \ad[1](P)\rightarrow T[1]P/G \rightarrow T[1]\Sigma \rightarrow 0~,
\end{equation}
where $\ad[1](P)$ is the grade-shifted vector bundle associated to $P$ by the adjoint action. A connection is then a splitting of this sequence as vector bundles, i.e.\ a map of graded manifolds (or, equivalently here, a bundle map) $a:T[1]\Sigma\rightarrow T[1]P/G$. If this splitting is a morphism of Lie algebras, then the connection is flat. 

Usually, the Courant (Lie 2-)algebroid is identified with a categorified Atiyah Lie algebroid; however, the complete picture is still unknown, as far as we are aware.

\subsection{General higher gauge theory}

We now generalize the picture of the previous section to arbitrary symplectic $L_\infty$-alge\-broids. We start again from the N$Q$-manifold $(T[1]\Sigma,\dd_\Sigma)$, assuming that $\Sigma$ is contractible. The gauge symmetries together with a potential $\sigma$-model target space are encoded in a metric $L_\infty$-algebroid, which we regard again as a symplectic N$Q$-manifold $(\CM,\omega,Q_\CM)$. We choose coordinates $Z^A=(y^a,\zeta^\alpha,\ldots)$ on $\CM=\CM_0\leftarrow \CM_1\leftarrow \ldots$. The {\em connective structure} (or {\em higher connection}) is encoded in a morphism of graded manifolds
\begin{equation}
 a:T[1]\Sigma\ \longrightarrow\ \CM~,
\end{equation}
which is not necessarily a morphism of N$Q$-manifolds. The map $a$
encodes a function $a_0$ on $\Sigma$ with values in $\CM_0$, which can
be regarded as a field in a (non-linear) sigma model. Moreover, $a$
contains $\CM_i$-valued $i$-forms $a_i$ over $\Sigma$ for all $i\leq \dim \Sigma$. Together, they contain the connective structure of an underlying higher principal bundle. The local kinematical data of higher gauge theory is therefore encoded in the map $a$.

The curvature of the connective structure, as well as the covariant derivative of the sigma model field $a_0$ are the failure of $a$ to be a morphism of N$Q$-manifolds. To develop this statement a little further, we consider the diagram
\begin{equation}\label{eq:4-diag}
  \xymatrixcolsep{5pc}
  \myxymatrix{
  T[1](T[1]\Sigma) \ar@{->}[r]^{a_*} & T[1]\CM  \\
  T[1]\Sigma\ar@{->}[u]^{\dd_\Sigma} \ar@{->}[r]^{a} & \CM\ar@{->}[u]^{Q_\CM}
  }
\end{equation}
where the homological vector fields $\dd_\Sigma$ and $Q_\CM$ are regarded as sections. Since both maps $a_*\circ \dd_\Sigma$ and $Q_\CM\circ a$ end in the same fiber over $\CM$, the map
\begin{equation}\label{eq:dga-curvature}
 f:=a_*\circ \dd_\Sigma-Q_\CM\circ a
\end{equation}
is well-defined and together with the trivial projection $\pi:T[1]\CM\rightarrow \CM$, $f$ yields a lift of $a$:
\begin{equation}
  \xymatrixcolsep{5pc}
  \myxymatrix{
  & T[1]\CM \ar@{->}[d]^{\pi} \\
  T[1]\Sigma \ar@{->}[ur]^{f} \ar@{->}[r]^{a} & \CM
  }
\end{equation}
We can now regard elements of $\Omega^\bullet(\CM)$ as functions on $T[1]\CM$ and pull them back along $f$. For $h\in \CC^\infty(\CM)$ in particular, we have 
\begin{equation}\label{eq:f-prop-1}
 f^*(\pi^*h)=a^*(h)\eand f^*(\dd_\CM h)=\left(\dd_\Sigma\circ a^*-a^*\circ Q_\CM\right)(h)~,
\end{equation}
where $\dd_\CM$ is the exterior derivative on $\CM$ and we used $Q_\CM^*\dd_\CM h=Q_\CM h$. Finally, because of the properties of the pullback of functions, we have 
\begin{equation}\label{eq:f-prop-2}
 f^*(\alpha\beta)=f^*(\alpha)f^*(\beta)
\end{equation}
for $\alpha,\beta\in \CC^\infty(T[1]\CM)$. Endowing $T[1]\CM$ with the homological vector field $\hat{Q}=\dd_{\CM}+\CL_{Q_\CM}$ with $\CL_{Q_\CM}:=\iota_{Q_\CM} \dd_\CM -\dd_\CM \iota_{Q_\CM}$ turns $f$ into a morphism of N$Q$-manifolds:
\begin{equation}\label{eq:Q-morphism}
  (\dd_\Sigma\circ f^*-f^*\circ \hat{Q})(\pi^*h)=0\eand (\dd_\Sigma\circ f^*-f^*\circ \hat{Q})(\dd_\CM h)=0~,
\end{equation}
which follows from equations \eqref{eq:f-prop-1}. Altogether, the covariant derivative of $a_0$ and the curvatures of the connective structure encoded in the $a_i$, for $i>0$, are encoded in the morphism of N$Q$-manifolds $f$.

Note that by enlarging the picture to $T[1]\CM$, we made the
transition from the Cheval\-ley-Eilenberg algebra (the functions on
$\CM$) to the Weil algebra (the functions on $T[1]\CM$), cf.\ section
\ref{ssec:Weil_algebra}. A morphism of N$Q$-manifolds $f:T[1]\Sigma\rightarrow T[1]\CM$ contains in particular a morphism of differential graded algebras (dga-morphism for short)  $f^*:\Walg(\CM)\rightarrow
\Walg(\Sigma)=\CC^\infty(T[1]\Sigma)\cong \Omega^\bullet(\Sigma)$. The connective structure is flat if and only if $f^*$ factors through the Chevalley-Eilenberg algebra $\CEalg(\CM)$. That is, in this case we have an $a^*$ such that
\begin{equation}
  \xymatrixcolsep{5pc}
  \myxymatrix{
  & \Walg(\CM) \ar@{->}[ld]_{f^*} \ar@{->}[d]^{\pi_\Walg} \\
  \Omega^\bullet(\Sigma)  & \ar@{->}[l]^{a^*} \CEalg(\CM)
  }
\end{equation}
is commutative.

As in the case of ordinary gauge theory, higher gauge transformations are again given by flat homotopies $\hat a:T[1](\Sigma\times[0,1])\rightarrow \CM$ between morphisms of graded manifolds $a,\tilde a:T[1]\Sigma\rightarrow \CM$. Alternatively, we can regard them as concordances between dga-mor\-phisms $f^*, \tilde f^*: \Walg(\CM)\rightarrow \CC^\infty(T[1]\Sigma)$ \cite{Sati:0801.3480}. 

Characteristic classes of the higher principal bundles endowed with the higher connective structure are given by pullbacks along $f$ of invariant polynomials on $T[1]\CM$ \cite{Kotov:2007nr}. For example, consider the symplectic form on $\CM$, given by $\omega=\tfrac12\dd Z^A \wedge \omega_{AB} \dd Z^B$ in some local Darboux coordinates $Z^A$. We regard $\omega$ as a function on $T[1]\CM$ and find its pullback along $f$ to be
\begin{equation}
 f^*(\tfrac12\dd Z^A\wedge \omega_{AB}\dd Z^B)=\tfrac12f^*(\dd Z^A)\wedge \omega_{AB}f^*(\dd Z^B)=\tfrac12F^A\wedge \omega_{AB}F^B~,
\end{equation}
where $F$ is the curvature of the connective structure. 

In the case of ordinary gauge theory with $\CM=\frg[1]$, the invariant polynomials are of the form $p=\frac{1}{d!}p_{A_1\ldots A_d}\dd Z^{A_1}\wedge \ldots \dd Z^{A_d}$. The corresponding pullbacks reproduce in particular all the Chern characters and the pullback of the symplectic form $f^*(\omega)$ is simply the second Chern class.

\subsection{Fake curvatures and dga-morphisms}

One of the goals of higher gauge theory is to describe the parallel transport of extended objects along submanifolds. To guarantee that this parallel transport is invariant under reparameterizations of these submanifolds, consistency conditions have to be imposed. These consistency conditions correspond to the vanishing of all so-called {\em fake curvatures}, see e.g.\ \cite{Baez:2004in}.

Vanishing of the fake curvature arises very naturally in twistor descriptions of higher gauge theory \cite{Saemann:2012uq,Saemann:2013pca,Jurco:2014mva}. It also renders the gauge transformation of the curvature covariant. The latter point is important in a potential discussion of the maximally superconformal field theory, or (2,0)-theory, in six dimensions. This theory contains a self-dual 3-form curvature, and imposing a self-duality condition is only gauge invariant for vanishing fake curvature. 

When regarding a connective structure as a dga-morphism
$f^*:\Walg(\CM)\rightarrow \CC^\infty(T[1]\Sigma)$, the fake curvatures
are precisely all but the highest form-degree element of the
corresponding curvatures. That is, vanishing of the fake curvatures means that the map $a$ fails to be a
morphism of N$Q$-manifolds only in its highest degree component. We
will comment more on this point when discussing examples. 

\subsection{Higher Chern-Simons actions and AKSZ \texorpdfstring{$\sigma$}{sigma}-models}\label{sec:actions-aksz-sigma}

Let us now go beyond the kinematical data and specify dynamical information via an action principle. A particularly interesting class of models can be obtained from the AKSZ formalism, cf.\ \cite{Alexandrov:1995kv,Roytenberg:2006qz,Kotov:2010wr,Fiorenza:2011jr}. The AKSZ formalism is a very general technique for constructing action functionals within the Batalin-Vilkovisky formalism. Here, we are interested in merely the resulting classical theories. These are generalizations of the Chern-Simons action functional in the sense that the corresponding Euler-Lagrange equations lead to completely flat connective structures. That is, we are looking for actions that force the morphism of graded manifolds $a:T[1]\Sigma\rightarrow M$ to be a morphism of N$Q$-manifolds. Accordingly, gauge transformations will turn out to be 2-morphisms of N$Q$-manifolds, which can be identified with homotopies between N$Q$-morphisms.

Consider a symplectic N$Q$-manifold of degree $n$, $(\CM,\omega,Q)$, as well as a morphism of graded manifolds $a:T[1]\Sigma\rightarrow \CM$, where $\Sigma$ is $n+1$-dimensional. Let $\chi$ be a Chern-Simons element, via which the Hamiltonian $\CS$ of $Q$ is transgressed to the invariant polynomial corresponding to the symplectic form. Then the functional corresponding to the pullback of $\chi$,
\begin{equation}\label{eq:action_AKSZ}
 S_{\rm AKSZ}=\int_\Sigma f^*(\chi)~,
\end{equation}
is the (classical) AKSZ action. For Poisson Lie algebroids, this construction yields Poisson sigma-models \cite{Bojowald:0406445}, and for Courant algebroids we obtain Courant sigma models \cite{Roytenberg:2006qz}. 

As a first example relevant to us, we reconstruct ordinary Chern-Simons theory in this way. Let $\Sigma$ be a three-dimensional manifold and $\frg$ a Lie algebra. We regard the Lie algebra as a symplectic N$Q$-manifold of degree 2 $(\frg[1],\dd_{\rm CE})$ with $\omega=\tfrac12\dd \xi^\alpha\wedge \omega_{\alpha\beta}\dd \xi^\beta$ in some suitable coordinates $\xi^\alpha$. The Chern-Simons element in the relevant transgression is
\begin{equation}
 \chi=\tfrac12\left(\xi^\alpha \omega_{\alpha\beta}\dd_\frg \xi^\beta-\tfrac{1}{3!}\omega_{\alpha\delta}f^\delta_{\beta\gamma}\xi^\alpha\xi^\beta\xi^\gamma\right)~,
\end{equation}
cf.\ section \ref{ssec:Weil_algebra}, and the corresponding AKSZ action functional is the ordinary Chern-Simons action functional of a connection on a trivial principal bundle over $\Sigma$:
\begin{equation}
 S_{\rm AKSZ}=\int_\Sigma f^*(\chi)=\tfrac12\int_\Sigma \left((A,F)-\tfrac{1}{6}(A,[A,A])\right)=\tfrac12\int_\Sigma(A,\dd_\Sigma A+\tfrac13[A,A])~,
\end{equation}
where we used the facts that $\omega_{\alpha\beta}$ encodes the Killing form and $f^*(\dd_\frg\xi^\alpha)=F^\alpha$.

Next, let us consider the first categorification in the form of four-dimensional higher Chern-Simons theory. Here, we start from a four-dimensional manifold $\Sigma$ and let $(\frg[1],\dd_{\rm CE})$ be the symplectic N$Q$-manifold of degree 3 given by $T^*[3]V[1]$, where $V$ is some vector space. This N$Q$-manifold was discussed in detail in section \ref{ssec:metric_L_infty_algebroids}. We will again use coordinates $\xi^\alpha$ on $V[1]$ and $p_\alpha$ in the fibers of $T^*[3]V[1]$. The symplectic form reads as $\omega=\dd \xi^\alpha\wedge \dd p_\alpha$ and the Hamiltonian $\CS$ of the homological vector field $Q$ given in \eqref{eq:Q_ex_1} reads as 
\begin{equation}
 \CS=\tfrac{1}{4}\iota_\eps\iota_Q \omega=-
  \tfrac{1}{4!} h_{\beta\gamma\delta\alpha}\xi^\alpha\xi^\beta\xi^\gamma\xi^\delta-\tfrac{1}{2}g^\gamma_{\beta\alpha}\xi^\alpha\xi^\beta p_\gamma -
  \tfrac{1}{2}e^{\alpha\beta}p_\alpha p_\beta~.
\end{equation}
The Chern-Simons element witnessing the transgression between $\omega$ and $\CS$ is given by
\begin{equation}
 \chi=\tfrac{1}{3}\left(\iota_\eps \omega+\CS\right)=\tfrac{1}{3}\big((\dd_\frg p_\alpha\xi^\alpha-2p_\alpha\dd_\frg\xi^\alpha)+\CS\big)~,
\end{equation}
which yields the AKSZ action functional
\begin{equation}\label{eq:AKSZ_4d}
 \begin{aligned}
  S_{\rm AKSZ}&=\int_\Sigma f^*(\chi)\\
  &=-\int_\Sigma \left[ \big( -B, \dd A+\tfrac{1}{2}\mu_2(A,A)
    -\tfrac{1}{2} \mu_1(B)\big) + \tfrac{1}{4!}\big( A, \mu_3(A,A,A)\big)\right]~.
 \end{aligned}
\end{equation}
The field content in this action functional is given by a local connective structure on a principal 2-bundle over $\Sigma$. That is, we have a 1-form potential $A=f^*(\xi)\in \Omega^1(\Sigma,V[1])$ together with a 2-form potential $-B=f^*(p)\in\Omega^2(\Sigma,V[2])$. The additional minus sign in front of $B$ is inserted for convenience, to match conventions most commonly used in the literature. Varying the action \eqref{eq:AKSZ_4d}, we obtain the following equations:
\begin{equation}\label{eq:27}
  \begin{aligned}
   0&=\dd A + \tfrac{1}{2}\mu_2(A,A)-\mu_1(B)=\mathcal{F}~,\\
   0&=\dd B +\mu_2(A,B)+\tfrac{1}{6}\mu_3(A,A,A)=\mathcal{H}~,
  \end{aligned}
\end{equation}
which evidently describe a flat connective structure on a principal 2-bundle with structure Lie 2-algebra $V\rightarrow W$. This motivates the identification of the AKSZ action functionals with higher Chern-Simons theories.

Note that for a strict Lie 2-algebras with $\mu_3=0$, we obtain the usual $BF$-theory. In particular, if $\mu_1=\id:V\cong W\rightarrow W$, we arrive at a $(BF+BB)$-type action, cf.\ the examples in \cite{Baez:2010ya}. 

The above construction can be readily extended to any metric Lie
$p$-algebra, or equivalently any N$Q$-manifold, or even
any metric Lie $p$-algebroid, or its corresponding N$Q$
manifold. The $p$-algebra action is the usual
\begin{equation}
  S_{\text{CS}}=\tfrac{1}{n}\int_\Sigma\left(|B|a^*(Z^B)\omega_{BA}\left(\dd_\Sigma a^*(Z^A)+a^*(Q^A)\right)+a^*(\mathcal{S})\right)~,
\end{equation}
where we indicate by $\dd_\Sigma$ the exterior derivative on the $(p+2)$-dimensional space-time
manifold $\Sigma$. We now wish to make contact with the traditional description of $L_\infty$-algebras in terms of higher brackets. For this, we move again the grading $|A|$ from the (weight-shifted) coordinates $Z^A$, to the (unshifted) generators of the
$p$-algebra $\hat\tau_A$, cf.\ equation \eqref{NPQ-mfd and n-algebras, brackets}. Coordinates now have degree
zero, while $|\hat\tau_A|=|A|-1$ and we group them all into one
$Z=\sum_AZ^A\hat\tau_A$ of mixed degree. Recall that in these conventions, the
products $\mu_k$ carry degree $k-2$. We have that
\begin{equation}
  \mathcal{S}=\sum_{k=1}^{p+1}\frac{(-1)^{\sigma_k}}{(k+1)!}m^B_{C_1\cdots C_k}  Z^{C_1}\cdots
  Z^{C_k}\omega_{BA}Z^A~.
\end{equation}
Let us assign the nomenclature
\begin{equation}
  a^*(Z^A)=: \phi^A~,\qquad
  a^*(Z)=\phi^A\hat\tau_A=:\phi_{\mu_1\cdots\mu_r}^A\hat\tau_A\otimes\dd
  x^{\mu_1}\wedge\cdots\wedge\dd x^{\mu_r}~,
\end{equation}
where $r=|A|\leq p$ is the weight of the $Z^A$ coordinate. This way,
taking the difference of the $\NN$-grading and the de Rham degree, each
component of $\phi$ has total degree $(-1)$ and thus the correct
behavior under permutations inside the multi-brackets.\\
We can exchange the structure constants
$m^B_{C_1\cdots C_k}$ for the brackets $\mu_k$:
\begin{equation}
  \mu_k(\phi,\ldots,\phi)=\mu_k(\phi^{C_1}\hat\tau_{C_1},\ldots,
  \phi^{C_k}\hat\tau_{C_k})=\phi^{C_1}\cdots\phi^{C_k}\otimes m^B_{C_1\cdots C_k}\hat\tau_B~.
\end{equation}
The action then becomes
\begin{equation}\label{n-algebra chern-simons action}
  S_{\text{CS}}=\int_\Sigma\left(\langle
  \phi,\dd_\Sigma\phi\rangle+\sum_{k=1}^{p+1}\frac{(-1)^{\sigma_k}}{(k+1)!}\langle \mu_k(\phi,\ldots,\phi),\phi\rangle\right)~,
\end{equation}
and the equations of motion can therefore be written as 
\begin{equation}
  \frac{\overleftarrow{\partial}S_{\text{CS}}}{\partial \phi^A}=-2\dd_M\phi^B\omega_{BA}+\sum_{k=1}^{p+1}\frac{(-1)^{\sigma_k}}{k!}\mu_k(\phi,\ldots,\phi)^B\omega_{BA}=0~,
\end{equation}
where we readily recognize the homotopy Maurer-Cartan equations\footnote{cf.\ section \ref{ssec:L_infty-models}} appearing, separated degree by degree.
We shall return to a more general discussion of equations of motions of dimensionally reduced AKSZ actions in section \ref{sec:L_infty_models}.

\section{Supersymmetric gauge theory with \texorpdfstring{N$Q$}{NQ}-manifolds}

Recall that the IKKT model is obtained by dimensionally reducing ten-dimensional maximally supersymmetric Yang-Mills theory, and we expect a similar relation between relevant $L_\infty$-algebra models and higher gauge theories. It is therefore important to understand both supersymmetric field theories as well as their dimensional reductions within the framework presented in the previous sections.

\subsection{Maximally supersymmetric field theories}\label{ssec:max_susy_field_theories}

Ordinary field theories on Minkowski space that do not involve gravity can have maximally 16 real supercharges. The representation theory of the corresponding supersymmetry algebras in the various dimensions give then rise to the possible field contents, cf.\ \cite{Seiberg:1997ax}. Here, we shall be interested in maximally supersymmetric Yang-Mills theories and superconformal field theories in six dimensions.

All maximally supersymmetric Yang-Mills theories on flat $d$-dimensional Minkowski space $0\leq d<10$, are obtained by dimensionally reducing $\CN=1$ super Yang-Mills theory on ten-dimensional Minkowski space $\FR^{1,9}$ \cite{Brink:1976bc}. To describe this theory, consider a principal fiber bundle over $\FR^{1,9}$ with structure group $\sG$ and connection $\nabla$ together with a Majorana-Weyl spinor $\psi$ taking values in the adjoint representation of the Lie algebra $\frg$ of $\sG$. The Majorana-Weyl spinor has 16 real components and satisfies
\begin{equation}
 \psi=C\bar{\psi}^T\eand \psi=+\Gamma\psi~,
\end{equation}
where $C$ is the charge conjugation operator and $\Gamma=\di \Gamma_0\ldots \Gamma_9$, being the product of all generators $\Gamma_\mu$ of the Clifford algebra $\CC\ell(\FR^{1,9})$, has chiral spinors as eigenspinors.

The action of $\CN=1$ super Yang-Mills theory reads as
\begin{equation}\label{eq:action_SYM_N1_10d}
 S=\int_{\FR^{1,9}} \left(-\tfrac{1}{4}(F,\star F)+\tfrac{\di}{2}{\rm vol}(\bar \psi,\nablas \psi)\right)~.
\end{equation}
Here, $F$ is the curvature of the connection $\nabla$, inner products on $\frg$ and its adjoint representation are denoted by $(\cdot,\cdot)$, ${\rm vol}$ is the volume form on $\FR^{1,9}$ and $\nablas$ is the usual Dirac operator. In the following, we will choose the standard basis on $\FR^{1,9}$ and work with components
\begin{equation}
 \nabla=\dd x^\mu\left(\der{x^\mu}+A_\mu\right)=\dd x^\mu(\dpar_\mu+A_\mu)\eand F=\tfrac{1}{2} F_{\mu\nu}\dd x^\mu\wedge \dd x^\nu~.
\end{equation}
The equations of motion resulting from varying \eqref{eq:action_SYM_N1_10d} read as
\begin{equation}\label{eq:eom_SYM_N1_10d}
 \nabla^\mu F_{\mu\nu}=-\tfrac12 \bar \psi \Gamma_\nu \psi\eand \nablas \psi=\Gamma^\mu\nabla_\mu \psi=0~. 
\end{equation}
They are invariant under the supersymmetry transformations
\begin{equation}\label{eq:SUSY_SYM_N1_10d}
 \delta A_\mu=\di \epsb \Gamma_\mu \psi\eand \delta \psi=\Sigma_{\mu\nu} F^{\mu\nu}\eps~,
\end{equation}
where $\Sigma_{\mu\nu}:=\tfrac14(\Gamma_\mu\Gamma_\nu-\Gamma_\nu\Gamma_\mu)$ and $\eps$ is the Majorana-Weyl spinor parameterizing the 16 supercharges. The action \eqref{eq:action_SYM_N1_10d} is also invariant under the supersymmetry transformations \eqref{eq:SUSY_SYM_N1_10d} up to terms that vanish on-shell, i.e.\ after imposing the equations of motion.

A manifestly supersymmetric formulation of the action of Yang-Mills theory with more than four real supercharges is difficult at best, and not readily available in the maximal case of 16 supercharges. This is due to the fact that for more than four real supercharges, we have to keep track of an infinite number of auxiliary fields to close the supersymmetry algebra off-shell. At the level of equations of motion, however, a superspace formulation does exist and we will briefly review the details in the following. We start from the superspace $\FR^{1,9|16}$ with bosonic coordinates $x^\mu$, $\mu=0,\ldots 9$ and fermionic coordinates $\theta^\alpha$, the latter forming a 16-component Majorana-Weyl spinor in ten dimensions. On $\FR^{1,9|16}$, we introduce the fermionic derivatives
\begin{equation}\label{eq:fermionic_derivatives}
 D_\alpha:=\der{\theta^\alpha}+\Gamma^\mu_{\alpha\beta}\theta^\beta \der{x^\mu}~,
\end{equation}
which satisfy the algebra
\begin{equation}
 \{D_\alpha,D_\beta\}=2\Gamma^\mu_{\alpha\beta}\der{x^\mu}~.
\end{equation}
The connection and curvature now have components along the fermionic directions
\begin{equation}
 \begin{aligned}
    \CA&=\CA_\mu \dd x^\mu+\CA_\alpha \dd \theta^\alpha~,\\
    \CF&=\tfrac12 \CF_{\mu\nu}\dd x^\mu\wedge \dd x^\nu+\CF_{\mu\alpha}\dd x^\mu\wedge \dd \theta^\alpha+\tfrac12\CF_{\alpha\beta}\dd \theta^\alpha\wedge \dd \theta^\beta~,
 \end{aligned}
\end{equation}
where the components of the curvature along vectors $X,Y$ on $\FR^{1,9|16}$ are given as usual by 
\begin{equation}
 \CF(X,Y):=\nabla_X Y-\nabla_Y X-\nabla_{[X,Y]}~.
\end{equation}
Moreover, the connection, its curvature and the spinor $\Psi$ are now superfields and depend on both $x^\mu$ and $\theta^\alpha$.

Interestingly, one can show that supergauge equivalence classes of solutions to the super Yang-Mills equations on $\FR^{1,9|16}$ given in terms of the superfields $\CA$ and $\Psi$,
\begin{equation}\label{eq:4.10}
 \nabla^\mu \CF_{\mu\nu}=-\tfrac12 \bar \Psi \Gamma_\nu \Psi\eand \nablas \Psi=\Gamma^\mu\nabla_\mu \Psi=0~,
\end{equation}
are in one-to-one correspondence to gauge equivalence classes of solutions to the ordinary super Yang-Mills equations \eqref{eq:eom_SYM_N1_10d} on $\FR^{1,9}$ \cite{Harnad:1985bc}.

An additional advantage of the superfield formulation is the fact that the supergauge equivalence classes of solutions to the super Yang-Mills equations on $\FR^{1,9|16}$ given in terms of the superfields can be equivalently described as superconnections which satisfy the following constraint equation \cite{Harnad:1985bc}
\begin{equation}\label{eq:constraint_SYM_N1_10}
 \{\nabla_\alpha,\nabla_\beta\}=2\Gamma^\mu_{\alpha\beta} \nabla_\mu~.
\end{equation}
Here, the spinor superfield $\Psi$ is identified with a curvature component according to
\begin{equation}
 \Psi^\alpha=\tfrac{1}{10}\Gamma^{\mu\alpha\beta}\CF_{\mu\beta}~,
\end{equation}
and the Bianchi identities imply the superfield equations of motion \eqref{eq:4.10}. We can state that solutions to the ten-dimensional super Yang-Mills equations are equivalent to superconnections on $\FR^{1,9|16}$ which are partially flat.

As mentioned above, maximally supersymmetric Yang-Mills theories in lower dimensions are obtained from the ten-dimensional theory by a simple dimensional reduction. The components of the gauge potential along the reduced directions turn into the scalar fields of the theory, while the Majorana-Weyl spinor $\psi$ in ten dimensions breaks up into corresponding spinors in lower dimensions.

A similar formulation exists of the maximally supersymmetric conformal field theory in six dimension, which was obtained by a twistor description in \cite{Saemann:2012uq,Saemann:2013pca,Jurco:2014mva}. The relevant superspace here is $\FR^{1,5|16}$ with coordinates $x^\mu$ and $\eta_I^\alpha$, where $I=1,\ldots,4$ is the R-symmetry index, and $\alpha=1,\ldots 4$ are (chiral) spinor indices. The R-symmetry group is $\sSp(2)$, regarded as elements of $\sSU(4)$ leaving invariant an antisymmetric $4\times 4$-matrix $\Omega^{IJ}$. The fermionic derivative reads as
\begin{equation}
 D^I_\alpha:=\der{\eta^\alpha_I}-2\Omega^{IJ}\sigma^\mu_{\alpha\beta}\eta^\beta_J\der{x^\mu}~,
\end{equation}
and satisfies
\begin{equation}
 \{D^I_\alpha,D^J_\beta\}=-2\Omega^{IJ}\sigma^\mu_{\alpha\beta}\der{x^\mu}~.
\end{equation}
Here, $\sigma^\mu_{\alpha\beta}$ are the analogue of the Pauli sigma matrices in six dimensions and we also introduce $\bar{\sigma}^{\mu\alpha\beta}=\tfrac{1}{4!}\eps^{\alpha\beta\gamma\delta}\sigma^\mu_{\gamma\delta}$. Consider now a principal 2-bundle over $\FR^{1,5|16}$ with connective structure. The latter is described by global 1- and 2-forms, taking values in the subspaces of degree $0$ and $1$ of a 2-term $L_\infty$-algebra. The resulting curvatures 
\begin{equation}
 \begin{aligned}
    \CF&:=\dd A+\tfrac12\mu_2(A,A)-\mu_1(B)~,\\
    \CH&:=\dd B+\mu_2(A,B)+\tfrac{1}{3!}\mu_3(A,A,A)
 \end{aligned}
\end{equation}
have to satisfy the following constraint equations:
\begin{equation}
 \begin{aligned}
    \CF&=0~,\\
    \sigma^\mu\bar \sigma^\nu \sigma^\kappa \CH_{\mu\nu\kappa}&=0~,\\
    (\sigma^\mu\bar{\sigma}^{\nu})_\beta{}^\gamma \CH_{\mu\nu}{}_\alpha^I&=\delta^\gamma_\alpha \Psi^I_\beta-\tfrac14\delta_\beta^\gamma \Psi_\alpha^I~,\\
    \sigma^\mu_{\gamma\delta}\CH_{\mu}{}_{\alpha\beta}^{IJ}&= \eps_{\alpha\beta\gamma\delta}\Phi^{IJ}~,\\
    \CH_{\alpha\beta\gamma}^{IJK}&=0~.
 \end{aligned}
\end{equation}
The second equation above amounts to $\CH=\star \CH$ on the purely bosonic part of $\CH$. The other equations identify scalar and spinor fields belonging to the $\CN=(2,0)$ supermultiplet in six dimensions. The Bianchi identities will result in field equations together with $\Omega_{IJ}\Phi^{IJ}=0$. We will come back to this theory in section \ref{sec:L_infty_models}.

\subsection{Dimensional reduction}\label{ssec:dimensional_reduction}

If we want to describe maximally supersymmetric Yang-Mills theories in less than 10 dimensions using morphisms between N$Q$-manifolds, we clearly need a mechanism of dimensionally reducing the theory. By dimensional reduction, physicists usually mean the process of compactifying one or several dimensions of space-time to a torus and letting its radii shrink to zero. In this process, the fields acquire discrete momentum modes in the compactified directions and the energy of all modes except for the constant or zero-mode diverges as the radii shrink. This justifies to neglect these modes in the lower-dimensional considerations.

Let us be slightly more general in the following. We start from our usual setup of a contractible manifold $\Sigma$ and regard its parity-shifted tangent bundle $T[1]\Sigma$ as an N$Q$-manifold, where $Q=\dd_\Sigma$. Consider now a set of integrable vector fields generating a subgroup $\sT$ of the isometries on our space-time manifold $\Sigma$ along the directions we want to reduce. Imposing the condition that all relevant fields and equations are invariant under this group $\sT$ amounts to dimensionally reducing our equations. Our reduction should be compatible with the $Q$-structure on $\Sigma$, which will be guaranteed if the homological vector field $Q$ commutes with the integrable vector fields generating $\sT$.

As a simple but relevant example, consider the reduction of $\Sigma=\FR^{1,9}$ with coordinates $x^M$, $M=0,\ldots,9$ to $\FR^{1,3}$ with coordinates $x^\mu$, $\mu=0,\ldots,3$. Let $\sT\cong\FR^{6}$ be the abelian group of translations in $\FR^{1,9}$ generated by the vector fields $\der{x^i}$, $i=4,\ldots,9$. These clearly commute with $Q_{\Sigma}=\xi^M\der{x^M}$, as $\xi^M$ is constant. Consider now a connection on a principal bundle over $\Sigma$ encoded in a morphism of graded manifolds $a:T[1]\Sigma\rightarrow \frg[1]$. Imposing the condition that the connective structure $a$ is invariant under $\sT$ restricts $a$ to a morphism of graded manifolds $a_{\rm red}:T[1]\FR^{1,3}\times\FR^{6}[1]\rightarrow \frg[1]$. Correspondingly, gauge transformations in the form of homotopies are restricted in the evident way. It is now easy to see that $a_{\rm red}$ encodes a gauge potential on $\FR^{1,3}$ with values in the Lie algebra $\frg$ together with six scalar fields $\phi^i$ in the adjoint representation of $\frg$. The lifted morphism of N$Q$-manifolds $f$ of \eqref{eq:dga-curvature} contains both the field strength of the gauge potential as well as the covariant derivatives of the scalar fields.

In our above reduction, matter fields in the adjoint representation arise from extending the tangent bundle $T[1]\FR^{1,3}$ by additional directions. Ideally, however, we would like matter fields to appear from extending $\frg[1]$ to an action Lie algebroid instead. Let us comment on this point in the following.

Given an action $\acton$ of a Lie group $\sG$ on some manifold $X$, we have a corresponding action Lie groupoid $X//\sG$ which is $\sG\times X\rightrightarrows X$, where $\sfs(g,x)=x$ and $\sft(g,x)=g\acton x$ for all $(g,x)\in \sG\times X$. We are interested in the case where $X=\frg^n=\sLie(\sG)^n$ and $\acton$ is the adjoint (and diagonal) action of $\sG$ on $\frg^n$. The corresponding action Lie algebroid $\sLie(X//\sG)$ is the trivial bundle $E=\frg^n\times \frg\rightarrow \frg^n$, where the anchor is implicit in the action of $\frg$ on $\frg^n$. The corresponding N$Q$-manifold is given by $E[1]$ and the differential $Q$ is simply the sum of the anchor map and the Chevalley-Eilenberg differential of $\frg$.

A morphism of graded manifolds $\phi:\FR^6[1]\rightarrow \frg[1]$ is necessarily linear and therefore characterized by a matrix $\phi^\alpha_i$, $\alpha=1,\ldots \dim(\frg)$, $i=1,\ldots 6$. This matrix also encodes a map $\tilde \phi:*\rightarrow \frg^6$. Therefore the reduced morphism of graded manifolds $a_{\rm red}:T[1]\FR^{1,3}\times\FR^{6}[1]\rightarrow \frg[1]$ can be regarded as a morphism of graded manifolds $a'_{\rm red}:T[1]\FR^{1,3}\rightarrow \frg^6\times \frg[1]$, where $\frg^6\times \frg[1]$ is the total space of the action Lie algebroid discussed above for $n=6$. Gauge symmetries act evidently appropriately, and the morphism of N$Q$-manifolds $f$ of \eqref{eq:dga-curvature} still encodes both the curvature of the gauge potential as well as the covariant derivative of the scalar field.

\subsection{Supersymmetric gauge theories using \texorpdfstring{N$Q$}{NQ}-manifolds}\label{ssec:SUSY_FT_via_NQ}

We now have everything at our disposal to implement supersymmetric field theories in the N$Q$-manifold picture. A na\"ive approach would be to introduce all component fields by hand. It is, however, much more natural to use the superfield formalism introduced in section \ref{ssec:max_susy_field_theories}. As stated before, we will not be able to write down manifestly supersymmetric actions, as off-shell formulations for supersymmetric theories with more than four real supercharges are cumbersome at best and usually not available at all.

We start from the superspace $\Sigma=\FR^{1,9|16}$ together with its parity shifted tangent bundle $T[1]\Sigma$. Note that we are considering a bigrading: the shift happens in a different $\NN$-grading from the $\RZ_2$-grading of the superspace $\Sigma$. Recall that an object with bidegree $(p,q)$, $p\in \NN$ and $q\in \RZ_2$ is even if $p+q$ is even and odd otherwise. We choose local coordinates $x^\mu$, $\mu=0,\ldots,9$, and $\theta^\alpha$, $\alpha=1,\ldots,16$ on the base of $T[1]\Sigma$, which are even and odd, respectively, as well as coordinates $\xi^\mu$ and $t^\alpha$ in the fibers of $T[1]\Sigma$, which are odd and even, respectively. The homological vector field reads therefore as
\begin{equation}
 Q_{\Sigma}=\xi^\mu\der{x^\mu}+t^\alpha\der{\theta^\alpha}~.
\end{equation}
A morphism of graded manifolds (which we assume to merely respect the $\NN$-grading) $a:T[1]\Sigma\rightarrow \frg[1]$ now encodes a gauge potential on $\FR^{1,9}$ together with a spinor field on the same space in the adjoint representation of $\frg[1]$, which form the field content of $\CN=1$ super Yang-Mills theory in ten dimensions. The equations of motion are now readily imposed in the form of constraint equation \eqref{eq:constraint_SYM_N1_10}. We demand that
\begin{equation}\label{eq:NQ_SYM_constraints_10d}
f^*(\dd_{\frg} w^a)\left(D_\alpha,D_\beta\right)=0~,
\end{equation}
where $w^a$ are coordinates on $\frg[1]$ and $D_\alpha$ are the fermionic derivatives introduced in \eqref{eq:fermionic_derivatives}.

To obtain lower dimensional maximally supersymmetric Yang-Mills theories, we have to dimensionally reduce the above picture. Let us consider the case of $\CN=4$ super Yang-Mills theory in somewhat more detail. Here, we reduce by factoring out the abelian group generated by the integrable vector fields $\der{x^i}$, $i=4,\ldots,9$. This reduces $\Sigma$ to $\Sigma_{\rm red}=T[1]\FR^{1,3|16}\times \FR^6[1]$. The space $\FR^{1,3|16}$ is now coordinatized by $x^\mu$, $\mu=0,\ldots,3$ and $\theta^{i\alpha}$, $\bar \theta^\ald_i$, where $i=1,\ldots,4$ and $\alpha,\ald=1,2$. Again, we shift from a morphism of $\NN$-graded manifolds $a_{\rm red}:\Sigma_{\rm red}\rightarrow \frg[1]$ to a morphism $a'_{\rm red}:T[1]\FR^{1,3|16}\rightarrow \sLie(\frg^6//\sG)$, where $\sLie(\frg^6//\sG)$ is the action Lie algebroid of the diagonal adjoint action of $\sG$ on $\frg^6$. The equations of motion are now imposed by demanding a dimensionally reduced form of \eqref{eq:NQ_SYM_constraints_10d}:
\begin{equation}
\begin{aligned}
f^*(\dd_E w^a)\left(D_{(\alpha i},D_{\beta)j}\right)&=0~,\\
f^*(\dd_E w^a)\left(\bar D_{(\ald}^i,\bar D_{\bed)}^j\right)&=0~,\\
f^*(\dd_E w^a)\left(D_{\alpha i},\bar D_{\bed}^j\right)&=0~,
\end{aligned}
\end{equation}
where $D_{\alpha i}$ and $\bar D^i_\ald$ are dimensional reductions of the fermionic derivatives $D_\alpha$ of \eqref{eq:fermionic_derivatives} and $w^a$ are coordinates in the total space of $E:=\sLie(\frg^6//\sG)$.

As a side remark, note that one can similarly describe $\CN=4$ self-dual Yang-Mills theory in four dimensions \cite{Devchand:1995vj}. This theory has the same supermultiplet as $\CN=4$ super Yang-Mills theory, but the anti-selfdual field strength is replaced by an auxiliary field of helicity -1. Its equations of motion comprise $F=\star F$ together with equations for the remaining fields in the multiplet. They are invariant under the chiral part of the $\CN=4$ supersymmetry algebra (i.e.\ under 8 supercharges). In this case, the equations of motion can be encoded on the N$Q$-manifold $T[1]\FR^{1,3|8}\rightarrow \sLie(\frg^6//\sG)$ arising from $T[1]\FR^{1,3|8}\times \FR^6[1]$ as above as follows:
\begin{equation}
 f^*(\dd_E w^a)(\dpar_{\alpha(\ald},\dpar_{\beta\bed)})=0~,~~~f^*(\dd_E w^a)(D^i_{(\ald},\dpar_{\beta\bed)})=0~,~~~
 f^*(\dd_E w^a)(D^i_{(\ald},D^j_{\bed)})=0~,
\end{equation}
where $\dpar_{\alpha\ald}$ is the coordinate derivative in spinor notation: $\dpar_{\alpha\ald}\sim \sigma_{\alpha\ald}^\mu\der{x^\mu}$.

Fully analogously, we obtain a description of the six-dimensional superconformal field theory. Here, we start from the supermanifold $\Sigma=\FR^{1,5|16}$ with coordinates as introduced in section \ref{ssec:max_susy_field_theories}. The N$Q$-manifold describing the gauge structure is a 2-term $L_\infty$-algebra $V[2]\rightarrow W[1]$ with local coordinates $v_m$ and $w^\alpha$. A morphism of $\NN$-graded manifolds $a:T[1]\Sigma\rightarrow V[2]\times W[1]$ captures the local connective structure of a principal 2-bundle over $\Sigma$, as well as spinor and scalar superpartners, cf.\ \cite{Saemann:2012uq,Saemann:2013pca,Jurco:2014mva}. The relevant constraint equations comprise
\begin{equation}
  f^*(\dd_{W[1]} w^a)=0\eand  \sigma^\mu\bar \sigma^\nu \sigma^\kappa f^*(\dd_{V[2]} v_m)\left(\der{x^\mu},\der{x^\nu},\der{x^\kappa}\right)=0~.
\end{equation}
The first equation yields a vanishing fake curvature (i.e.\ the 2-form part $\CF$ of the curvature $f$ vanishes), while the second equation implies self-duality of the 3-form part $\CH$ of $f$ as well as the corresponding supersymmetric matter field equations.

\section{Nambu-Poisson and multisymplectic manifolds}

\subsection{The IKKT model and emergent geometry}

The Ishibashi-Kawai-Kitazawa-Tsuchiya (IKKT) model \cite{Ishibashi:1996xs,Aoki:1998bq} is the 0-dimen\-sional field theory or matrix model
\begin{equation}\label{eq:action_IKKT}
 S_{\rm IKKT}=\alpha\tr\left(-\tfrac{1}{4}\,[X_\mu,X_\nu]^2-\tfrac{1}{2}\,\psib\Gamma^\mu[X_\mu,\psi]+\beta \unit\right)~,
\end{equation}
where $X_\mu$ are scalar fields obtained by Kaluza-Klein reduction of a gauge potential on $\FR^{1,9}$ and $\psi$ are their superpartners, both taking values in $\au(\CH)$, where $\CH$ is a Hilbert space with countable dimension. Often, we are interested in deformations of this model arising from turning on background fluxes. A prominent such deformation is
\begin{equation}\label{eq:action_IKKT_def}
 S_{\rm def}=S_{\rm IKKT}+\tr\left( -\tfrac{1}{2}\sum_\mu m^2_{1,\mu}X_\mu X_\mu+\tfrac{\di}{2}m_2 \psib \psi+c_{\mu\nu\kappa}X^\mu X^\nu X^\kappa\right)~.
\end{equation}

The IKKT model can be obtained by regularizing the Schild action of the type IIB superstring with target $\FR^{1,9}$. In this action, sigma model fields, Poisson brackets and integrals appear, which can be regularized by matrices, Lie brackets and traces. This process amounts to replacing the Poisson manifold $\FR^{1,9}$ with a quantization in terms of ten-dimensional Moyal space. Alternatively, dimensionally reducing maximally supersymmetric Yang-Mills theory on $\FR^{1,9}$ to zero dimensions also yields the IKKT model. 

This model has been suggested as a background independent formulation
of string theory. For a truly background independent formulation,
ten-dimensional Minkowski space still features uncomfortably
prominently in the IKKT model. On the other hand, this is
due to the  IKKT model being the most symmetric zero-dimensional model
with 32 supercharges (including $\kappa$-symmetry). 

\subsection{Nambu-Poisson structures}

The IKKT model is particularly attractive as quantized spacetimes ``emerge'' as solutions of the model. Moreover, small fluctuations of these solutions can be interpreted as gauge field theories on the corresponding quantized spacetimes \cite{Aoki:1999vr}. One might now go a step further and ask to what extent there is a dynamics of these quantized spacetimes, see \cite{Yang:2006hj,Steinacker:1003.4134} and references therein.

A strong limitation is given by the fact that geometrically quantized
manifolds arising as solutions to the IKKT model are automatically 
K{\"a}hler. To get arbitrary manifolds, one is naturally led to generalizations of the IKKT model involving Nambu-Poisson structures \cite{Arnlind:2011rz,Arnlind:1312.5454}. This is very much in agreement with string theory considerations, which show that the generalized Schild action for higher branes can be reformulated in terms of Nambu-Poisson brackets, too \cite{Park:2008qe}. Also, recall that the Bagger-Lambert-Gustavsson M2-brane model \cite{Bagger:2007jr,Gustavsson:2007vu} is based on a 3-Lie algebra\footnote{not to be confused with a Lie 3-algebra}, which can be regarded as a quantized Nambu-Poisson structure.

A {\em Nambu-Poisson structure} of rank $n$ on a manifold $M$ is a multivector field $\pi\in \Gamma(\wedge^n TM)$ such that the {Nambu-Poisson bracket}
\begin{equation}
 \{f_1,\ldots,f_n\}=\pi(\dd f_1,\ldots,\dd f_n)~,~~~f_i\in \CC^\infty(M)~,
\end{equation}
satisfies the {\em fundamental identity}
\begin{equation}\label{eq:fundamental_identity}
\begin{aligned}
 \{f_1,\ldots,f_{n-1},\{g_1,\ldots,g_n\}\}=&\\
 \{\{f_1,\ldots,f_{n-1},&g_1\},\ldots,g_n\}\}+\ldots+ \{g_1,\ldots,\{f_1,\ldots,f_{n-1},g_n\}\}~,
\end{aligned}
\end{equation}
cf.\ \cite{Nambu:1973qe,Takhtajan:1993vr}. One can show \cite{springerlink:10.1007/BF00400143} that in a neighborhood of points $p\in M$, where $\pi(p)\neq 0$, there exist local coordinates $x^1,\ldots,x^n,x^{n+1},\ldots,x^d$, $d=\dim(M)$, such that
\begin{equation}
 \pi=\der{x^1}\wedge \ldots\wedge \der{x^n}~.
\end{equation}
Conversely, a multivector field that is locally of this form yields a Nambu-Poisson structure on $M$. 

Note that Nambu-Poisson structures form $n$-Lie algebras in the sense of \cite{Filippov:1985aa} and therefore naturally yield {\em strict} $p$-term $L_\infty$-algebras with $p=n-1$, cf.\ the discussion in \cite{Palmer:2012ya}. Here, however, we are interested in semistrict $p$-term $L_\infty$-algebras and do not follow this connection any further.

Given a Nambu-Poisson structure of rank $n$ on some manifold $M$, we can associate to every antisymmetric $p$-tuple of functions $f_\bullet=f_1\wedge \ldots\wedge f_p\in \wedge^p\CC^\infty(M)$ a Hamiltonian vector field $X_{f_\bullet}$ via
\begin{equation}\label{eq:Ham_vec_multifuncs}
 X_{f_\bullet}(g):=\{f_1,\ldots,f_p,g\}
\end{equation}
for all $g\in\CC^\infty(M)$ \cite{springerlink:10.1007/BF00400143}. One can readily verify that the fundamental identity \eqref{eq:fundamental_identity} is equivalent to demanding that $\CL_{X_{f_\bullet}}\pi=0$, cf.\ e.g.\ \cite{Nakanishi1999}.

In order to generalize the IKKT model to an extended model which is a regularization of a sigma-model involving Nambu-Poisson brackets, we need a quantization of Nambu-Poisson structures. This is a very subtle problem and for a detailed discussion from our point of view, we refer to \cite{DeBellis:2010pf} and references therein. Here, let us merely recall that in geometric quantization, the Hamiltonian vector fields are to be lifted to an algebra of quantum observables. One might therefore be led to interpreting the Lie algebra of the above defined Hamiltonian vector fields as the algebra of quantum observables on quantized Nambu-Poisson manifolds. As we shall explore now, there is more structure on the space of {\em multifunctions} $\wedge^{p}_\FC\CC^\infty(M)$ that underlies these vector fields, and it seems more natural to quantize these structures instead.

Using the Hamiltonian vector field \eqref{eq:Ham_vec_multifuncs}, we can define a product on $\wedge^{p}_\FC\CC^\infty(M)$ \cite{Takhtajan:1994aa}:
\begin{equation}
\begin{aligned}
 f_\bullet*g_\bullet:=&\sum_{k=1}^{p} g_1\wedge \ldots \wedge X_{f_\bullet}(g_k)\wedge \ldots \wedge g_{p}\\
 =&\sum_{k=1}^{p} g_1\wedge \ldots \wedge \{f_1,\ldots,f_{p},g_k\}\wedge \ldots \wedge g_{p}~.
\end{aligned}
\end{equation}
This product satisfies the Jacobi identity
\begin{equation}\label{eq:star_jacobi}
 f_\bullet*(g_\bullet*h_\bullet)=(f_\bullet*g_\bullet)*h_\bullet+g_\bullet*(f_\bullet*h_\bullet)~,
\end{equation}
and thus we have a Leibniz algebra structure on $\wedge^{p}_\FC\CC^\infty(M)$. This Leibniz algebra induces a Lie algebra structure on the set of Hamiltonian vector fields since the fundamental identity directly translates into the relation
\begin{equation}
 [X_{f_\bullet},X_{g_\bullet}]=X_{f_\bullet*g_\bullet}~,
\end{equation}
where $[-,-]$ denotes the Lie bracket of vector fields.

We can now antisymmetrize the product to obtain a generalization of a Lie structure \cite{Nakanishi:1998:499-510,Vaisman:1999aa}. We define
\begin{equation}
 \pi_2(f_\bullet,g_\bullet)=\tfrac12(f_\bullet*g_\bullet-g_\bullet*f_\bullet)~,
\end{equation}
which also yields
\begin{equation}\label{eq:preservation}
 [X_{f_\bullet},X_{g_\bullet}]=X_{\pi_2(f_\bullet,g_\bullet)}~.
\end{equation}
The antisymmetrization of $*$ to $\pi_2$ clearly broke the Jacobi identity \eqref{eq:star_jacobi}, and because of \eqref{eq:preservation}, the failure to satisfy the Jacobi identity is an element of the kernel of $X_-:\wedge^{p}_\FC\CC^\infty(M)\rightarrow \frX(M)$. Therefore, the usual discussion in the literature proceeds to define a set of {\em Casimir multifunctions} $\CCC(M)$ which form the kernel of the map $X_-$. The map $X_-$ then becomes a Lie algebra isomorphism between $\wedge^{p}_\FC\CC^\infty(M)/\CCC(M)$ and the Lie algebra of Hamiltonian vector fields.

Knowing about the relevance of $L_\infty$-algebras and the controlled violation of the Jacobi identity, we do not follow this route but work with the space $\wedge^{p}_\FC\CC^\infty(M)$ together with the products $*$ and $\pi_2$. Indeed, we can define the following brackets, turning $\CCC(M)\embd\linebreak \wedge^{p}_\FC\CC^\infty(M)$ into a 2-term $L_\infty$-structure: $\pi_1:\CCC(M)\embd \wedge^{p}_\FC\CC^\infty(M)$ is just the embedding, $\pi_2$ is defined on $\CCC(M)\wedge (\wedge^{p}_\FC\CC^\infty(M))$ and $(\wedge^{p}_\FC\CC^\infty(M))\wedge (\wedge^{p}_\FC\CC^\infty(M))$ as the above $\pi_2$ and $\pi_3$ is defined implicitly via 
\begin{equation}
\begin{aligned}
 \pi_2(f_\bullet,\pi_2(g_\bullet,h_\bullet))+\pi_2(g_\bullet,\pi_2(h_\bullet,f_\bullet))+\pi_2(h_\bullet,\pi_2(f_\bullet,g_\bullet))=\pi_1(\pi_3(f_\bullet,g_\bullet,h_\bullet))~.
\end{aligned}
\end{equation}
This 2-term $L_\infty$-algebra is rather uninteresting and instead of dwelling further on it, let us come to more relevant structures.

\subsection{Heisenberg Lie \texorpdfstring{$p$}{p}-algebra}

A more interesting picture emerges after restricting to multifunctions $\CH^{p}(M)$ within\linebreak $\wedge^{p}_\FC \CC^\infty(M)$, which consist of linear and constant functions in some local coordinates $x^i$, $i=1,\ldots, d$, $d=\dim(M)$. This is sensible, as this set of operators forms a complete set of classical observables. Knowing the value of these observables, we know the value of all classical observables (i.e.\ functions on $M$). We further assume that these coordinates are the ones in which the Nambu-structure decomposes, i.e.\
\begin{equation}
 \pi=\der{x^1}\wedge \ldots \wedge \der{x^n}~.
\end{equation}
We immediately extend the picture to the complex
\begin{equation}
 \FR\longembd \CH^1(M)\xrightarrow{~\mu_1~}\CH^2(M)\xrightarrow{~\mu_1~}\ldots \xrightarrow{~\mu_1~}\CH^{p}(M)~~~\big(\xrightarrow{~X_-} \frX(M)\big)~,
\end{equation}
where $\mu_1(f_\bullet):=1\wedge f_\bullet$, with $\mu_1\circ \mu_1=0$ obvious. As indicated by the parentheses, we can regard this complex as a resolution of $\frX(M)$. Using theorem 7 of \cite{Barnich:1997ij} together with the second remark after that theorem, we conclude that this complex can be endowed with a $p$-term $L_\infty$-structure. We call the resulting $L_\infty$-algebra the {\em Heisenberg Lie $p$-algebra} $\CH^\bullet(M)$ of the Nambu-Poisson manifold $M$.

In fact, one can readily define the explicit brackets and these will be useful in the subsequent discussion. We introduce the basis elements
\begin{equation}
\begin{aligned}
 \vartheta_{i_1\ldots i_m}&:=\eps_{i_1\ldots i_mj_1\ldots j_{n-m}}x^{j_1}\wedge \ldots \wedge x^{j_{n-m}}\in \CH^{n-m}(M)~,\\
 \zeta_{i_1\ldots i_m}&:=\eps_{i_1\ldots i_mj_1\ldots j_{n-m}}1\wedge x^{j_1}\wedge \ldots \wedge x^{j_{n-m}}\in \CH^{n-m+1}(M)~,
\end{aligned}
\end{equation}
where again $p=n-1$. The product $\pi_2$ on $\wedge^{p}\CC^\infty(M)$ restricts to the following products on $\CH^{p}(M)$:
\begin{equation}
 \mu_2(\vartheta_i,\vartheta_j)=(-1)^{p}\eps_{ijk_1\ldots k_{p-1}}1\wedge x^{k_1}\wedge \ldots \wedge x^{k_{p-1}}~,~~~
 \mu_2(\vartheta_i,\zeta_{jk})=\mu_2(\zeta_{ij},\zeta_{k\ell})=0~.
\end{equation}
We put all other $\mu_k$ for $k\geq 2$ to zero except for
\begin{equation}
 \mu_k(\vartheta_{i_1},\vartheta_{i_2},\ldots,\vartheta_{i_k}):=(-1)^{\sigma_k}
 \varepsilon_{i_1\cdots i_k i_{k+1}\cdots i_n}1\wedge
 x^{i_{k+1}}\wedge\cdots\wedge x^{i_n}~, 
\end{equation}
for $\vartheta_l\in\CH^{p}(M)$. 
It is trivial to check that the higher homotopy relations are satisfied. 

Since this $L_\infty$-algebra is finite-dimensional, we can formulate it also in terms of an N$Q$-manifolds. The underlying graded vector space is
\begin{equation}
 0 \longleftarrow \FR^{\binom{d+1}{p}}[1]\longleftarrow \FR^{\binom{d+1}{p-1}}[2]\longleftarrow \ldots \longleftarrow \FR^{\binom{d+1}{1}}[p] \longleftarrow 0
\end{equation}
with local coordinates $(\check \vartheta^{i_1\ldots i_j})$ and $(\check \zeta^{i_1\ldots i_j})$, and the homological vector field $Q$ is given by
\begin{equation}
 Q=\sum_{k=2}^{p+1}(-1)^{\binom{k+1}{2}}\check \vartheta^{i_1}\check \vartheta^{i_2}\ldots \check\vartheta^{i_k} \der{\check \zeta^{i_1i_2\ldots i_k}}+\sum_{j=2}^{p+1}\check \vartheta^{i_1\ldots i_j}\der{\check \zeta^{i_1\ldots i_j}}~.
\end{equation}
Here, the first and the second sum correspond to $\mu_k$ for $k\geq 2$ and $\mu_1$, respectively. This Heisenberg Lie $p$-algebra is, however, just a part of a much more interesting structure, which is obtained by switching to a dual picture in terms of multisymplectic geometry.

\subsection{Multisymplectic manifolds}

Just as Poisson structures that provide a non-degenerate map from $T^*_xM$ to $T_xM$ yield symplectic structures, Nambu-Poisson structures that provide non-degenerate maps from $\wedge^{p} T^*_xM$ to $T_xM$ yield multisymplectic structures. Recall that a {\em multisymplectic form of degree $n$} or a {\em $p$-plectic form} on a manifold $M$ is a closed form $\varpi\in \Omega^n(M)$ such that $\iota_X \varpi=0$ is equivalent to $X=0$ for all $X\in \frX(M)$. Note that while a symplectic form on $M$ always gives rise to Poisson structures on $\CC^\infty(M)$, multisymplectic forms do not necessarily give rise to Nambu-Poisson structures. If, however, the multisymplectic form $\varpi$ arises from a volume form on $M$ regarded as a map $\wedge^{p} TM\rightarrow T^*M$, then the inverse map is given by a multivector field which encodes a Nambu-Poisson structure. Inversely, if a Nambu-Poisson structure $\pi$ is non-degenerate in the sense that its corresponding multivector field provides a non-degenerate map $\pi:\wedge^{p} T^*M\rightarrow T_xM$ for all $x\in M$, then its inverse is necessarily a volume form and therefore a multisymplectic form on $M$. This follows from the results of \cite{springerlink:10.1007/BF00400143}. In the following, we will be mostly interested in this non-degenerate case.

Given a manifold with a multisymplectic form $\varpi$ of degree $n=p+1$, we can define a set of Hamiltonian $(p-1)$-forms $\Omega^{p-1}_{\rm Ham}(M)$, which consists of those forms $\alpha$ for which
\begin{equation}
 \iota_{X_\alpha}\varpi=\dd \alpha
\end{equation}
for some $X_\alpha\in \frX(M)$. Consider now the case that $\varpi$ is the inverse of a Nambu-Poisson structure $\pi$. There is a natural map $\delta:\wedge^{p}_\FC\CC^\infty(M)\rightarrow \Omega^{p-1}(M)$ defined as
\begin{equation}
 \delta(f_1\wedge f_2\wedge \ldots \wedge f_{p}):=\frac{1}{p!}\eps^{i_0\,i_1\ldots i_{p}} f_{i_1}\dd f_{i_2}\wedge \ldots \wedge \dd f_{i_{p}}~.
\end{equation}
We then have the identity
\begin{equation}
 X^\pi_{f_\bullet}=X^\varpi_{\delta(f_\bullet)}~.
\end{equation}
That is, the Hamiltonian vector field for an element $f_\bullet\in\wedge^{p}_\FC\CC^\infty(M)$ computed with respect to the Nambu-Poisson structure $\pi$ agrees with that of $\delta(f_\bullet)$ computed with respect to the multisymplectic structure $\varpi$. This is clear because
\begin{equation}
 X^\pi_{f_\bullet}=\pi(\dd f_1,\ldots,\dd f_{p},-)\eand \iota_{X^\varpi_{\delta(f_\bullet)}}\varpi=\dd f_1\wedge \dd f_2\wedge \ldots \wedge \dd f_{p}~.
\end{equation}

The products $*$ and $\pi_2(-,-)$ we introduced above on $\wedge^{p}_\FC\CC^\infty(M)$ turn into products $*$ and $\pi_2(-,-)$ on the Hamiltonian $(p-1)$-forms as follows:
\begin{equation}
\begin{aligned}
 \delta(f_\bullet*g_\bullet)&=\delta(f_\bullet)*\delta(g_\bullet):=\CL_{X^\varpi_{\delta f_\bullet}}\delta(g_\bullet)~,\\
 \delta(\pi_2(f_\bullet,g_\bullet))&=\pi_2(\delta(f_\bullet),\delta(g_\bullet)):=-\iota_{X^\varpi_{\delta (f_\bullet)}}\iota_{X^\varpi_{\delta(g_\bullet)}}\varpi~.
\end{aligned}
\end{equation}
The products on Hamiltonian $p-1$-forms have been introduced before \cite{Cantrijn:1996aa} and in \cite{Baez:2008bu}, they were called the hemi-bracket and the semi-bracket, respectively. Altogether, we recovered the Lie algebra of Hamiltonian vector fields and the additional product structure also in the multisymplectic case.

Note that the above products $*$ and $\pi_2(-,-)$ on Hamiltonian $(p-1)$-forms can be defined on arbitrary multisymplectic manifolds: 
\begin{equation}
 \alpha * \beta:= \CL_{X_\alpha}\beta\eand \pi_2(\alpha,\beta):=-\iota_{X_\alpha}\iota_{X_\beta}\varpi
\end{equation}
for $\alpha,\beta\in \Omega^{p-1}_{\rm Ham}(M)$. The product $*$ satisfies the Leibniz rule, but it is not antisymmetric. The product $\pi_2$ is antisymmetric but does not satisfy the Jacobi identity. The latter product is interesting to us and it can be extended to an $n$-term $L_\infty$-algebra on
\begin{equation}\label{eq:sh-complex}
 \FR\ \longembd\ \CC^\infty(M)\xrightarrow{~\dd~}\Omega^1(M)\xrightarrow{~\dd~}\ldots \xrightarrow{~\dd~}\Omega^{n-3}(M)\xrightarrow{~\dd~}\Omega^{p-1}_{\rm Ham}(M)~,
\end{equation}
as shown in \cite{Rogers:2010nw} using results of \cite{Barnich:1997ij}. The brackets here are given by $\mu_1=\dd$ and the other brackets all vanish except for 
\begin{equation}
 \mu_k(\alpha_1,\ldots,\alpha_i)=(-1)^{\binom{k+1}{2}}\iota_{X_{\alpha_1}}\ldots\iota_{X_{\alpha_k}}\varpi~,~~~2\leq k\leq n~,
\end{equation}
for $\alpha_i\in \Omega^{p-1}_{\rm Ham}(M)$. We will refer to this $L_\infty$-algebra as the {\em strong homotopy Lie algebra of local observables}, or {\em shlalo} for short.

When we restrict ourselves to forms which are constant or linear in the coordinates and minimally complete this set to a closed Lie $p$-algebra with respect to the original brackets, we obtain the {\em Heisenberg Lie $p$-algebra} on the corresponding complexes. One readily sees that for multisymplectic manifolds whose multisymplectic form arises from a Nambu-Poisson structure, the restricted shlalo is isomorphic to the Heisenberg Lie $p$-algebra of the Nambu-Poisson structure.

As a nontrivial example of a 2-plectic manifold, consider the natural 2-plectic structure on a compact simple Lie group $\sG$ given by \cite{Cantrijn:1999aa}
\begin{equation}
 \varpi(X,Y,Z)=\langle X,[Y,Z]\rangle+\langle Y,[Z,X]\rangle+\langle Z,[X,Y]\rangle
\end{equation}
for $X,Y,Z\in \frX(\sG)$,
where $\langle -,-\rangle$ is the appropriately normalized Killing
form and $[-,-]$ is the Lie bracket on the Lie algebra of $\sG$. From
this 2-plectic structure, we obtain an infinite dimensional Lie
2-algebra. Moreover, the sub Lie 2-algebra of $\sG$-left-invariant one-forms is the string Lie 2-algebra of $\sG$ \cite{Baez:2009:aa}.

There is an interesting connection between symplectic N$Q$-manifolds and shlalos of multisymplectic manifolds. As shown in \cite{Roytenberg:1998vn}, Courant algebroids, i.e.\ symplectic N$Q$-manifolds concentrated in degrees $0$, $1$ and $2$, come with a natural 2-term $L_\infty$-algebra. In the special case of twisted Vinogradov algebroids, this 2-term $L_\infty$-algebra contains an $L_\infty$-subalgebra, which is isomorphic to the shlalo of a 2-plectic manifold \cite{Rogers:2010sc}. Later, this picture was extended, and the general statement is the following. Given a $p$-plectic manifold $(M,\varpi)$, one can construct the Vinogradov algebroid $T^*[p]T[1]M$ over $M$ twisted by $\varpi$, cf.\ \cite{Gruetzmann:2014ica}, which comes naturally with a $p$-term $L_\infty$-algebra \cite{Fiorenza:0601312,Getzler:1010.5859}. This $L_\infty$-algebra then contains an $L_\infty$-subalgebra which is isomorphic to the shlalo of the $p$-plectic manifold $M$ \cite{Ritter:2015ffa}. 

\subsection{Quantization and emergence of multisymplectic manifolds}

While a full procedure for quantizing multisymplectic manifolds has yet to be developed, it is not difficult to infer some of its desired properties as done in \cite{Ritter:2013wpa}. This will be sufficient for our subsequent discussion.

Essentially, classical quantization consists of a Lie algebra
homomorphism to first order in $\hbar$, which maps the Poisson algebra
of smooth functions on a Poisson manifold to a Lie algebra of quantum
observables. Here, $\hbar$ is a parameter that governs the
quantization. From our above discussion, it seems clear that the Poisson algebra of smooth functions should be replaced by the $p$-term $L_\infty$-algebras arising on Nambu-Poisson manifolds with non-degenerate Nambu-Poisson tensor of rank $n=p+1$. These form the shlalo of the corresponding $p$-plectic manifold. Therefore, a quantization is an $L_\infty$-algebra homomorphism to first order in $\hbar$ mapping this shlalo to an $L_\infty$-algebra of multisymplectic quantum observables. This approach is in line with that of \cite{Rogers:2011zc,Fiorenza:1304.6292,Fiorenza:2013kqa}.

There are now essentially two possibilities for defining homomorphisms of $p$-term $L_\infty$-algebras, see \cite{Ritter:2015ffa} for a very detailed discussion. We can simply regard them as graded vector spaces endowed with brackets and demand that a homomorphism is a chain map of the underlying graded vector spaces which commutes with all the brackets in the obvious way. On the other hand, one might want to regard $p$-term $L_\infty$-algebras as Lie $p$-algebras, and correspondingly as $p$-categories. Then a homomorphism should be a weak $p$-functor between the underlying categories. The latter is also very reasonable from regarding $L_\infty$-algebras as N$Q$-manifolds, because it corresponds to morphisms of N$Q$-manifolds.  In \cite{Ritter:2013wpa}, however, all interesting results were obtained from the former approach and we therefore restrict to chain maps in the following.

In classical quantization, the quantization homomorphism up to first order in $\hbar$ often restricts to an exact homomorphism of Lie algebras on the Heisenberg algebra. Moreover, we say that a K\"ahler manifold is a vacuum solution of a set of equations of motion if the generators of its Heisenberg algebra (i.e.\ the subalgebra of the Poisson algebra consisting of linear and constant functions) provide a non-trivial solution to these equations.

We expect to see the same in higher quantization. That is, the quantization homomorphism up to first order in $\hbar$ should restrict in the relevant cases to an exact homomorphism of $L_\infty$-algebras on the Heisenberg Lie $p$-algebra $\CH^\bullet(M)$. In particular, by saying that a multisymplectic manifold of rank $n$ is a vacuum solution of a set of equations of motion, we mean that these equations are non-trivially solved by the generators of the Heisenberg Lie $p$-subalgebra of the shlalo of the manifold.

As in the case of the IKKT model, we will often rely on embeddings of our multisymplectic manifolds into flat Euclidean space and use the Heisenberg algebra in these coordinates. For example, the fuzzy sphere arises as a solution of the IKKT model with a 3-form background field as a quantization of the Heisenberg algebra \cite{Iso:0101102,Kimura:0103192}
\begin{equation}
 \{x^i,x^j\}=\eps^{ijk}x^k~,
\end{equation}
where $x^i$ are Cartesian coordinates on $\FR^3$ with $|x|=1$, describing $S^2$ as being embedded into $\FR^3$. The corresponding picture for $S^3$ was given in \cite{Ritter:2013wpa} and below, we generalize this to arbitrary $S^n$, $n\geq 2$.

\subsection{Central Lie \texorpdfstring{$p$}{p}-algebra extensions of \texorpdfstring{$\mathfrak{so}(n+1)$}{so(n+1)}}\label{ssec:central_extension}

Consider the sphere $S^n$ embedded in $\FR^{n+1}$ as $|x|=1$. We would like to determine the corresponding Heisenberg Lie $p$-algebra $\CH^\bullet(S^n)$ in terms of the Cartesian coordinates $x^i$. All our equations below hold modulo $|x|^2=1$. The generators of the Heisenberg Lie $n$-algebra are 
\begin{equation}
\begin{aligned}
 &1~,&&x^\mu~,&&x^{\mu_1}x^{\mu_2}~,~~~\ldots~,~~~x^{\mu_1}\ldots x^{\mu_p}~,\\
 &&&\dd x^\mu~,&&x^{\mu_1}\dd x^{\mu_2}~,~~~\ldots~,~~x^{\mu_1}\ldots x^{\mu_{p-1}}\dd x^{\mu_p}~,\\
 &&&&&\dd x^{\mu_1}\wedge\dd x^{\mu_2}~,~~~\ldots~,~~x^{\mu_1}\ldots x^{\mu_{p-2}}\dd x^{\mu_{p-1}}\wedge \dd x^{\mu_p}~,\\
 &&&\ldots\\
 &&&&&\vartheta_{\mu\nu}:=\tfrac12 \eps_{\mu\nu \lambda_1\ldots \lambda_{p}}x^{\lambda_1}\dd x^{\lambda_2}\wedge \ldots \wedge \dd x^{\lambda_{p}}~.
\end{aligned}
\end{equation}
Note that all generators are central except for $\vartheta_{\mu\nu}$. To calculate the higher brackets, note that the volume form of $S^n$ reads as
\begin{equation}
 {\rm vol}_{S^n}=\tfrac{1}{n!}\eps_{\mu_1\ldots \mu_{n+1}}x^{\mu_1}\dd x^{\mu_2}\wedge \ldots \wedge \dd x^{\mu_{n+1}}=:\varpi~.
\end{equation}
The Hamiltonian vector fields
$X_{\vartheta_{\mu\nu}}=X^\kappa_{\vartheta_{\mu\nu}}\der{x^\kappa}$
should be in $T S^n$ and therefore $X^\kappa_{\vartheta_{\mu\nu}}
x^\kappa=0$. To determine $X_{\vartheta_{\mu\nu}}$, we use the
equation $\iota_{X_{\vartheta_{\mu\nu}}}\varpi=\dd \vartheta_{\mu\nu}$
in the form
\begin{equation}
 x^\kappa\wedge \dd x^\kappa \wedge (\iota_{X_{\vartheta_{\mu\nu}}}{\rm vol}_{S^n})=x^\kappa \wedge \dd x^\kappa\wedge \dd \vartheta_{\mu\nu}~.
\end{equation}
Using the usual identities for the totally antisymmetric tensors of $\mathfrak{so}(n+1)$, we find that
\begin{equation}
 X_{\vartheta_{\mu\nu}}=x^\mu\der{x^\nu}-x^\nu\der{x^\mu}~,
\end{equation}
and the Lie algebra of Hamiltonian vector fields is $\mathfrak{so}(n+1)$, as expected. The Heisenberg Lie $p$-algebra product $\pi_2$ reads as
\begin{equation}\label{eq:cat_of_so4}
 \pi_2(\vartheta_{\mu\nu},\vartheta_{\kappa\lambda})=\delta_{\nu\kappa}\vartheta_{\mu\lambda}-\delta_{\mu\kappa}\vartheta_{\nu\lambda}-\delta_{\nu\lambda}\vartheta_{\mu\kappa}+\delta_{\mu\lambda}\vartheta_{\nu\kappa}+\pi_1(R_{\mu\nu\kappa\lambda})~,
\end{equation}
where\footnote{Note that our paper \cite{Ritter:2013wpa} contained an typographical error, and the last two signs of $R_{\mu\nu\kappa\lambda}$ were interchanged in the discussion of $S^3$.} in the case $n=3$,
\begin{equation}
 R_{\mu\nu\kappa\lambda}=\tfrac{1}{4}\left(\eps_{\nu\kappa\lambda\rho}x^\rho x^\mu-\eps_{\mu\kappa\lambda\rho}x^\rho x^\nu+\eps_{\kappa\mu\nu\rho}x^\rho x^\lambda-\eps_{\lambda\mu\nu\rho}x^\rho x^\kappa\right)~.
\end{equation}
Altogether, we see that the Heisenberg Lie $p$-algebra of $S^n$ is a central extension of the Lie algebra $\mathfrak{so}(n+1)$ to a Lie $p$-algebra. 

This result matches an important expectation: In the case $n=3$, the resulting Lie 2-algebra agrees at the level of vector spaces with the 3-Lie algebra $A_4$ underlying the BLG model \cite{Bagger:2007jr,Gustavsson:2007vu}; see also \cite{Palmer:2012ya} and \cite{Palmer:2013ena} for connections of this model with higher gauge theory. 

\section{\texorpdfstring{$L_\infty$}{L-infinity}-algebra models}\label{sec:L_infty_models}

In our previous paper \cite{Ritter:2013wpa}, we distinguished between {\em homogeneous Lie 2-algebra models}, in which the fields were simply elements of a 2-term $L_\infty$-algebra and {\em inhomogeneous Lie 2-algebra models}, in which each field was an element of a homogeneously graded subspace of an $L_\infty$-algebra. The homogeneous models form a special case of the inhomogeneous ones and have a larger symmetry group. 

We will start by discussing homogeneous $L_\infty$-algebra models of Yang-Mills type, before continuing with the inhomogeneous $L_\infty$-algebra models arising from a dimensional reduction of the gauge part of the six-dimensional $(2,0)$-theory. We then show how dimensionally reduced higher Chern-Simons theories provide a unifying picture.

\subsection{Homogeneous \texorpdfstring{$L_\infty$}{L-infinity}-algebra models of Yang-Mills type}

As a simple example of a homogeneous $L_\infty$-algebra model, we
consider fields $X^I$, $I=1,\ldots,N$ taking values in a metric $L_\infty$-algebra $\sL$ together with the following obvious generalization of the IKKT action:
\begin{equation}\label{eq:homogeneous_action}
\begin{aligned}
 S=\tfrac14 \big(\mu_2(X^I,X^J),&\mu_2(X^I,X^J)\big)+\tfrac12m(X^I,X^I)\\&+\sum_{i\geq 3} \frac{(-1)^{\sigma_i}}{i!}c_{I_1\ldots I_i}\big(X^{I_1},\mu_{i-1}(X^{I_2},\ldots,X^{I_i})\big)~, 
\end{aligned}
\end{equation}
where the outer brackets denote the cyclic inner product, $m$ is a mass term and the $c_{I_1\ldots I_i}$ are constants, possibly arising from background fluxes in string or M-theory. The corresponding equations of motion are readily derived varying \eqref{eq:homogeneous_action} and using \eqref{eq:ax_cyclic_metric}:
\begin{equation}\label{eq:homogeneous_eom}
 \mu_2(X^I,\mu_2(X^J,X^I))+mX^J+\sum_{i\geq 2} (-1)^{\sigma_i}c_{JI_2\ldots I_i}\mu_{i-1}(X^{I_2},\ldots,X^{I_i})=0~.
\end{equation}

We now briefly generalize the special cases and their solutions discussed in \cite{Ritter:2013wpa} for homogeneous Lie 2-algebra models to $L_\infty$-algebra models.

If $m$ and all background fluxes $c_{I_1\ldots I_i}$ vanish, we
readily obtain a solution from the Heisenberg algebra $\CH^\bullet(\FR^n)$ of $\FR^n$, regarded as a $p$-plectic manifold with the multisymplectic form being the volume form. In Cartesian coordinates, we introduce the Hamiltonian $(p-1)$-forms
\begin{equation}\label{eq:Hamiltonian-forms}
 \vartheta_{i}=\tfrac{1}{p!}\eps_{ik_0\ldots k_{p-1}}x^{k_0}\dd x^{k_1}\wedge \ldots \wedge \dd x^{k_{p-1}}~,
\end{equation}
where $\eps_{i_1\ldots i_n}$ it the totally antisymmetric invariant tensor of $\mathfrak{so}(n)$. The corresponding Hamiltonian vector fields are $\der{x^i}\equiv\partial_i$ and 
\begin{equation}
 \pi_2(\vartheta_i,\vartheta_j)=-\iota_{\partial_i}\iota_{\partial_j}\varpi~.
\end{equation}
Because the latter is central in $\CH^\bullet(\FR^n)$, putting $X^i=\vartheta_i$ for $i=1,\ldots,n$ and $X^I=0$ otherwise, yields a solution to \eqref{eq:homogeneous_eom} for $m=c=0$. Note that for $n=2$, this is actually the usual way of obtaining the Moyal plane as a solution from the IKKT model. If $n$ is small enough compared to $N$, we can also consider Cartesian products of multisymplectic manifolds, just as Cartesian products of the Moyal plane solve the IKKT model.

There are now two complementary choices of parameters $m$ and $c$,
which lead to solutions of \eqref{eq:homogeneous_eom} built from the
Heisenberg Lie $p$-algebra of the spheres $S^n$, which we computed above in
section \ref{ssec:central_extension}. The field configuration $X^I$ we are interested in are given by
\begin{equation}\label{eq:homog_sphere_solutions}
 X^I=(\vartheta_{12},\ldots,\vartheta_{1(d+1)},\vartheta_{23},\ldots,\vartheta_{2(d+1)}\ldots,\vartheta_{d(d+1)},0,\ldots,0)~,
\end{equation}
for which
\begin{equation}
 \mu_2(X^I,\mu_2(X^J,X^I))=4(n+2)X^J~.
\end{equation}
Then \eqref{eq:homog_sphere_solutions} solves \eqref{eq:homogeneous_eom} with non-trivial mass
\begin{equation}
 m=-4(n+2)~.
\end{equation}
It is clear that one can similarly choose non-trivial $(n+1)$-form fluxes $c_{i_1\ldots i_{n+1}}$ or use a combination of mass and $(n+1)$-form fluxes to obtain the spherical solutions \eqref{eq:homog_sphere_solutions}.

Again, in the case $n=2$ of a symplectic form $\varpi$, this is the usual way of obtaining the fuzzy sphere as a solution of the mass or flux-deformed IKKT model. Cartesian products of spherical solutions can certainly be obtained, too, and they can also be generalized to products involving spheres and flat space.
  
\subsection{\texorpdfstring{$L_\infty$}{L-infinity}-algebra model of the \texorpdfstring{$(2,0)$}{(2,0)}-theory}\label{ssec:2-0-theory}

The maximally superconformal field theory in six dimensions, or $(2,0)$-theory for short, is in many ways the six-dimensional analogue of $\CN=4$ super Yang-Mills theory. Since the dimensional reduction of the latter yields the IKKT model, it is natural to ask what the dimensional reduction of the former might lead to.

Recall that the field content of the $(2,0)$-theory consists of a tensor multiplet containing a 2-form gauge potential with self-dual curvature 3-form. While even the existence of a classical description of the non-abelian $(2,0)$-theory is still an open question, we can make an obvious na\"ive guess. We simply consider higher gauge theory capturing the parallel transport of one-dimensional objects, which are supposed to correspond to the self-dual strings in the (2,0)-theory. Starting from a 2-term $L_\infty$-algebra $\frg=W\oplus V$, we have a potential one-form $A$ taking values in $W$ and a potential 2-form $B$ taking values in $V$. These come with curvatures
\begin{equation}
 \begin{aligned}
  \CF=\dd A+\tfrac12\mu_2(A,A)-\mu_1(B)\eand \CH=\dd B+\mu_2(A,B)+\tfrac{1}{3!}\mu_3(A,A,A)~,
 \end{aligned}
\end{equation}
which we demand to be fake-flat and self-dual, respectively:
\begin{equation}\label{eq:eom_2,0}
 \CF=0~,~~~\CH=\star \CH~.
\end{equation}
We then dimensionally reduce this theory to zero dimensions to obtain an inhomogeneous Lie 2-algebra model with equations of motion\footnote{We use weighted antisymmetrization: $\eps_{[ijk]}=\eps_{[ij]k}=\eps_{ijk}$, etc.}
\begin{equation}\label{eq:2,0-model}
\begin{aligned}
 \mu_2(X_M,X_N)-\mu_1(Y_{MN})&=0~,\\
 \mu_2(X_{[M},Y_{NK]})+\tfrac13\mu_3(X_M,X_N,X_K)&=\tfrac{1}{3!}\eps_{MNKPQR}\left(\mu_2(X^P,Y^{QR})+\tfrac13\mu_3(X^P,X^Q,X^R)\right)
\end{aligned}
\end{equation}
with $X_M\in W$ and $Y_{MN}\in V$. Note that the first equation implies that $\mu_2(X_M,X_N)$ is cohomologically trivial.

This model is solved by the Heisenberg Lie 2-algebra of $\FR^{1,2}\times \FR^{3}$ as follows. In terms of Cartesian coordinates $x^0,\ldots x^5$ on this space, this Heisenberg Lie 2-algebra is the direct sum of the Heisenberg Lie 2-algebras of $\FR^{1,2}$ and $\FR^3$. The latter two are in fact isomorphic and we define generators
\begin{equation}
\begin{aligned}
 &1~,~~~x^i~,~~~\dd x^i~,~~~&\zeta_i&:=\tfrac12 \eps_{ijk}x^j\dd x^k~,~~~i,j,k=0,1,2~,\\
 &1~,~~~x^a~,~~~\dd x^a~,~~~&\vartheta_a&:=\tfrac12 \eps_{abc}x^b\dd x^c~,~~~a,b,c=3,4,5~,
\end{aligned}
\end{equation}
We now put, for $i+3=a$,
\begin{equation}
 X_i=\zeta_i~,~~~X_{i+3}=\vartheta_a\eand
 Y_{ij}=\eps_{ijk}\dd x^k~,~~~Y_{ia}=Y_{aj}=0~,~~~Y_{ab}=\eps_{abc}\dd x^{c}~.
\end{equation}
This clearly solves \eqref{eq:2,0-model}. This is fully expected, since it is the categorified lift of the fact that noncommutative Minkowski space, as far as it can be sensibly defined, is a solution of the IKKT matrix model. 

\subsection{\texorpdfstring{$L_\infty$}{L-infinity}-algebra models from higher Chern-Simons theory}\label{ssec:L_infty-models}

Let us now come to a dimensional reduction of higher Chern-Simons theory. That is, we consider the AKSZ action \eqref{eq:action_AKSZ} and dimensionally reduce it to a point. For this, we can follow the procedure outline in section \ref{ssec:dimensional_reduction}. 

We start from an $n+1$-dimensional contractible manifold $\Sigma$ and a symplectic N$Q$-manifold $\CM$ of degree $n$ concentrated in positive degrees. Gauge configurations here are described by morphisms $f$ of N$Q$-manifolds from $T[1]\Sigma$ to $T[1]\CM$ with $f^*\circ \dd_\Walg=\dd_\Sigma\circ f^*$. After the dimensional reduction, this is reduced to a morphism of N$Q$-manifolds from $\FR^{n+1}[1]$, endowed with the trivial homological vector field, to $T[1]\CM$ with $f^*\circ (Q+\dd_\CM)=0$. The AKSZ action therefore reduces to possibly the most natural choice: a pullback to $\FR^{n+1}[1]$ of the Hamiltonian $\CS$ of the homological vector field $Q$ on $\CM$:
\begin{equation}
 S_{\rm AKSZ,0}= f^*(\CS)~~~\mbox{with}~~~\mathcal{S}=\sum_{k=1}^{n+1}\frac{(-1)^{\sigma_k}}{(k+1)!}m^B_{C_1\cdots C_k}  Z^{C_1}\cdots
  Z^{C_k}\omega_{BA}Z^A~,
\end{equation}
where $Z^A$ are coordinates on the vector space $\CM$. The pullback $f^*(Z)$ can be decomposed according to $\phi:=f^*(Z)=\sum_{k\geq 1} \sigma_k\varphi_{(k)}=A-B+\ldots$, where $\sigma_k$ is a factor inserted for convenience, cf.\ \eqref{eq:sigma_k}, and $\varphi_{(k)}$ take values in $\CM[-1]_k$ and are therefore homogeneous polynomials of degree $k$ in the Gra\ss mann coordinates parameterizing $\FR^{n+1}[1]$. In terms of $\phi$, the reduced AKSZ action reads now as follows:
\begin{equation}\label{eq:action_AKSZ_0}
 S_{\rm AKSZ,0}=-\frac{1}{n+1}\left(\phi,\sum_{k=1}^{n}\tfrac{(-1)^{\sigma_k}}{(k+1)!}\mu_k(\phi,\ldots,\phi)\right)~,
\end{equation}
where the inner product is a combination of the inner product on the original $L_\infty$-algebra and a projection of the product of the two entries onto the top component in terms of the Gra\ss mann coordinates.

It is now only natural to generalize the action \eqref{eq:action_AKSZ_0} to a {\em homotopy Maurer-Cartan action} $S_{\rm hMC}$ in which $\phi$ is an element of arbitrary degree in an arbitrary $L_\infty$-algebra. Such actions are relevant e.g.\ in the context of string field theory, cf.\ \cite{Zwiebach:1992ie}. The resulting equations of motion,
\begin{equation}
  \sum_i \frac{(-1)^{\sigma_k}}{k!}\mu_k(\phi,\ldots,\phi)\ =\ 0~,
\end{equation}
are known as the homotopy Maurer Cartan equations \cite{Lada:1992wc}. They are invariant under the gauge transformations \cite{Lada:1992wc}, see also \cite{Zwiebach:1992ie}:
\begin{equation}
  \phi\rightarrow \phi+\delta \phi\ewith \delta \phi\ =\ \sum_k \frac{(-1)^{\sigma_k-k}}{(k-1)!}\mu_k(\gamma,\phi,\ldots,\phi)~,
\end{equation}
where $\gamma$ is a degree 0 element of the $L_\infty$-algebra under consideration.

Because of the canonical choice of the action $S_{\rm hMC}$, the only input data into our model is the symplectic N$Q$-manifold, whose degree and dimensionality fixes completely the fields. This strict limitation of input data is a very appealing feature of this type of $L_\infty$-algebra model.

\subsection{Homotopy Maurer-Cartan equations and physically motivated models}

Let us now briefly connect the homotopy Maurer-Cartan equation to physically relevant models. There are essentially two ways of doing this. First, for gauge field theories whose equations amount to flatness conditions on a curvature along some subspace, we can develop twistor descriptions encoding the equations of motions in certain holomorphic vector bundles. These bundles are then described in terms of holomorphic Chern-Simons theory, which are clearly of homotopy Maurer-Cartan form. Examples for such field theories are the (super) Yang-Mills theories and self-dual Yang-Mills theories discussed in section \ref{ssec:SUSY_FT_via_NQ}, as well as the candidate (2,0)-theories mentioned in the same section.

Another possibility is to note that open and in particular closed string field theory \cite{Zwiebach:1992ie} has the homotopy Maurer-Cartan equation as equation of motion. In the derivation of Yang-Mills theory from string field theory, one has to rewrite the Yang-Mills equations in homotopy Maurer-Cartan form \cite{Zeitlin:2007vv,Zeitlin:2007yf}. Finally, one can readily guess a homotopy Maurer-Cartan form of both the Yang-Mills equations or e.g.\ the equations of motion of the BLG M2-brane models \cite{Bagger:2007jr,Gustavsson:2007vu}, as done in \cite{IuliuLazaroiu:2009wz}. We briefly demonstrate how this works in the case of the six-dimensional $(2,0)$-theory.

Consider a 2-term $L_\infty$-algebra $\frg=(V\stackrel{\mu_1}{\rightarrow}W)$ with products $\mu_k$, $1\leq k \leq 3$. We extend $\frg$ to a 3-term $L_\infty$-algebra $\tilde \frg=(F\stackrel{\mu_1}{\rightarrow} P\stackrel{\mu_1}{\rightarrow} C)$ of $\frg$-valued functions, potentials and curvatures as follows:
\begin{equation}
\begin{aligned}
 F&=\Omega^0(\FR^{1,5},W)~\oplus~\Omega^1(\FR^{1,5},V)~,\\
 P&=\Omega^1(\FR^{1,5},W)~\oplus~\Omega^2(\FR^{1,5},V)~,\\
 C&=\Omega^2(\FR^{1,5},W)~\oplus~\Omega^3(\FR^{1,5},V)~\oplus~\Omega^4(\FR^{1,5},W)~,
\end{aligned}
\end{equation}
where $F$, $P$ and $C$ are the elements of degree $0$, $1$ and $2$. All products $\tilde \mu_k$ on $\tilde \frg$ involving elements from $F$ or $C$ are defined to vanish, and on $P$ we define
\begin{equation}
\begin{aligned}
 \tilde \mu_1(p)&=\pm\dd p\mp*\dd p+\mu_1(p)-*\mu_1(p)~,\\
 \tilde \mu_2(p_1,p_2)&=\mu_2(p_1,p_2)-*\mu_2(p_1,p_2)~,\\
 \tilde \mu_3(p_1,p_2,p_3)&=\mu_3(p_1,p_2,p_3)-*\mu_3(p_1,p_2,p_3)~,
\end{aligned}
\end{equation}
where $*$ denotes the Hodge star on $\FR^{1,5}$ with respect to the canonical volume form and the signs in $\tilde \mu_1$ have to be chosen appropriately for forms $p$ of even and odd degree. The homotopy Maurer-Cartan equations on $\tilde g$ are equivalent to the pure gauge part of the (2,0)-theory \eqref{eq:eom_2,0} and read as
\begin{equation}
 \CF=*\CF=0~,~~~\CH=*\CH
\end{equation}
where $\CF=\dd A+\tfrac12\mu_2(A,A)-\mu_1(B)$ is the fake curvature and $\CH=\dd B+\mu_2(A,B)+\tfrac{1}{3!}\mu_3(A,A,A)$ is the 3-form curvature.

We conclude that with the right choice of $L_\infty$-algebra, the homotopy Maurer-Cartan equations can reproduce a wide range of physically relevant and interesting equations of motion.

\subsection{Multisymplectic solutions to \texorpdfstring{$L_\infty$}{L-infinity}-algebra Chern-Simons theory}
  
Let us now come to considering solutions to the zero-dimensional reduced higher Chern-Simons models we constructed previously. We expect in particular that $p$-plectic $\FR^n$, with the volume form as $p$-plectic structure,
always solves a twisted form of the homotopy Maurer-Cartan equations. As we had seen in
\cite{Ritter:2013wpa}, to obtain quantized $\FR^3$ as a solution to a
3-dimensional BF-theory, we needed to add a twist to the action, very
similar to the case of noncommutative Yang-Mills theory. Here, we
wish to incorporate the twist involving Lagrange multipliers, which originate
from singling out a direction of the unreduced AKSZ model. That is, we
would like to obtain quantum $\FR^n$ as a solution to the dimensional
reduction of $(n+1)$-dimensional higher Chern-Simons theory.

We start from a morphism of graded manifolds $\phi$ between the Gra\ss mann algebra $\FR^{n+1}[1]$ with coordinates $\psi^i$ and the Lie $p$-algebra $\frg[1]$ with basis $\tau_A$. We decompose the underlying fields according to 
\begin{equation}
 \phi=\sum_{k=1}^{p}\phi^A_{i_1\cdots i_k}\hat \tau_A \psi^{i_1}\cdots\psi^{i_k}~,
\end{equation}
where the degree of $\phi^A_{i_1\cdots i_k}\hat \tau_A$ (and thus that of $\hat \tau_A$) is $k-1$ in $\frg$. We pick the direction of $x^{n+1}$ and promote fields with components along this direction to Lagrange multipliers:
\begin{equation}
\phi_{n+1}:=\lambda~,\quad \phi_{i (n+1)}:=\lambda_i~,~~~\phi_{ij(n+1)}:=\lambda_{ij}~,~~~\ldots~~~\phi_{i_1i_2\cdots i_{k-1}(n+1)}:=\lambda_{i_1\cdots i_{k-1}}~. 
\end{equation}
In terms of these, the reduced AKSZ action reads as
\begin{equation}\label{0dim action 1}
 \begin{aligned}
  S_{\rm AKSZ,0}&=\eps^{j_1\cdots j_{n+1}}\left(
  \sum_{k=1}^{n}\tfrac{(-1)^{\sigma_k}}{(k+1)!}
  \left[\mu_k(\phi,\ldots,\phi)\right]_{j_1\cdots
    j_q},\tfrac{1}{(n+1-q)!}\phi_{j_{q+1}\cdots
                   j_{n+1}}\right)+t \\
&=\eps^{j_1\cdots j_{n}}\left(
  \sum_{k=1}^{n+1}\tfrac{(-1)^{\sigma_k}}{(k+1)!}
  \left[\mu_k(\phi,\ldots,\phi)\right]_{j_1\cdots
    j_q},\tfrac{1}{(n+1-q)!}\lambda_{j_{q+1}\cdots j_{n}}\right)+t~,
 \end{aligned}
\end{equation}
where $t$ is a twist to be specified later and by $\left[\mu_k(\phi,\ldots,\phi)\right]_{j_1\cdots j_q}$ we mean
  \begin{equation}
    \left[\mu_k(\phi,\ldots,\phi)\right]_{j_1\cdots
    j_q}=\sum_{l_1+\ldots+l_k=q}\mu_k(\tfrac{1}{l_1!}\phi_{i_1\cdots
    i_{l_1}},\ldots,\tfrac{1}{l_k!}\phi_{j_1\cdots j_{l_k}})~.
  \end{equation}
  
Varying this action with respect to $\phi_{j_1\cdots j_m}$, we obtain the equation of motion
\begin{equation}\label{eom 0-dim. n-alg. action}
 \begin{aligned}
    0=&\eps^{j_1\cdots j_{n+1}}
   \sum_{l_1,\ldots , l_k=1}^n\sum_{k=1}^{n}\tfrac{(-1)^{\sigma_k}}{(k+1)!}\tfrac{1}{l_1!}\cdots\tfrac{1}{l_k!}\tfrac{1}{(n+1-q)!}
  \Bigg[\delta_p^{n+1-q}\mu_k(\phi,\ldots,\phi)_{j_1\cdots
    j_q}\nonumber\\
&\hspace{+1cm}+\sum_{j=1}^k
  (-1)^{k-j+q(p-1)+l_j(l_{j+1}+\cdots+l_k)}\delta_p^{l_j}\mu_k(\phi,\ldots,
  \phi )_{j_1\cdots j_{n+1-l_j}}\Bigg]+t'~,
 \end{aligned}
\end{equation}
where $t'$ denotes potential contributions from the twist $t$. Let us start by considering the case $n=2$, for which the gauge $L_\infty$-algebra is an ordinary Lie algebra $\frg$, the field $\phi$ is given by $\phi=\phi_1\xi^1+\phi_2\xi^2+\lambda\xi^3$. The equation of motion for the $\phi_1$ and $\phi_2$ read as 
\begin{equation}
 [\phi_1,\phi_2]=t'_{12}~.
\end{equation}
For this equation to be solved by elements $\phi_i=x^i$ of the Heisenberg algebra $\{x^1,x^2\}=1$ on $\FR^2$, we clearly need to introduce a twist 
\begin{equation}
 t=\tfrac{(-1)^n}{(n+1)!}\left(\psi^1\ldots \psi^n,\lambda\right)~.
\end{equation}
This twist turns out to be the one that works also for all higher values of $n$. 

To find the solutions for higher $n$, we can proceed exactly as in the Lie $2$-algebra case discussed in \cite{Ritter:2013wpa}. In particular, we can consider $p$-plectic $\FR^{n}$, with $p=n-1$ and volume form $\varpi=\tfrac{1}{n!}\eps_{i_1\cdots i_{n}}\dd
x^{i_1}\wedge\cdots\wedge \dd x^{i_{n}}$ in terms of the usual Cartesian coordinates $x^i$ on $\FR^{n}$. On $(\FR^{n},\varpi)$, we construct the Heisenberg Lie $p$-algebra $\CH^\bullet$. In the basis $\vartheta_i$ of \eqref{eq:Hamiltonian-forms}, we have the non-trivial higher products
\begin{equation}
 \mu_1=\dd\eand \mu_k(\vartheta_{i_1},\ldots,\vartheta_{i_k})=(-1)^{\sigma_k} \iota_{\partial_{i_1}}\cdots\iota_{\partial_{i_k}}\varpi~.
\end{equation}
The equations of motion then read as 
\begin{equation}
 \begin{aligned}
  \tfrac{(-1)^{\sigma_k}}{n!}\mu_{n}(\phi_{i_1},\ldots,\phi_{i_{n}})&=\tfrac{(-1)^{\sigma_k}}{n!}\eps_{i_1\ldots i_{n}}~,~~~&\\
    \tfrac{(-1)^{\sigma_k}}{k!}\mu_k(\phi_{i_1},\ldots,\phi_{i_k})-\tfrac{2}{2!}\mu_1(\phi_{i_1\cdots i_k})&=0~,~~~&k<n~.
 \end{aligned}
\end{equation}
A solution is obtained by identifying the fields $\phi_i$ with the elements in a basis of $\CH^{p-1}$:
\begin{equation}\label{(n-1)-forms}
  \phi_i:=-\tfrac{1}{p!}\eps_{i\,j_1\cdots j_{p}}x^{j_1}\dd
  x^{j_2}\wedge\cdots\wedge\dd x^{j_{p}}~,
\end{equation}
and other fields according to 
\begin{equation}\label{(n-l)-forms}
  \phi_{i_1\cdots i_l}=(-1)^{\sigma_l}~\frac{1}{(n+1)!}~\binom{p+1}{l}~\eps_{i_1\cdots
                                           i_{p+1}}~x^{i_{l+1}}\dd
                                         x^{i_{l+2}}\wedge\cdots\wedge\dd
                                         x ^{i_{p+1}}~.
\end{equation}

Altogether, we can thus conclude that the expected quantization of multisymplectic $\FR^n$ comes with a Heisenberg Lie $p$-algebra which solves the twisted homotopy Maurer-Cartan equations. Note that the homotopy Maurer-Cartan equations are essentially invariant under $L_\infty$-algebra isomorphisms of the $L_\infty$-algebra under consideration. That is, even by choosing isomorphic Heisenberg algebras, we cannot remove the twist element.

\section*{Acknowledgements}
We would like to thank Thomas Strobl for very helpful discussions related to N$Q$-manifolds and their dimensional reduction. We are also grateful to Brano Jur\v co and Martin Wolf for useful conversations. We thank in particular Jim Stasheff for many helpful comments on a first draft of this paper. Our work was significantly simplified by the many helpful definitions and explanations found on the \href{http://ncatlab.org}{nLab} and we would like to thank all authors of this website. A first version of this paper was finished during the workshop ``Higher Structures in String Theory and Quantum Field Theory'' at the Erwin Schr\"odinger International Institute for Mathematical Physics in 2015 and we would like to thank the organizers and the institute for hospitality. The work of CS was partially supported by the Consolidated Grant ST/L000334/1 from the UK Science and Technology Facilities Council.

\bibliography{bigone}

\bibliographystyle{latexeu}

\end{document}